\documentclass{llncs}
\usepackage[a4paper,margin=25mm]{geometry} 
\usepackage{comment}

\usepackage{booktabs}
\usepackage{array}
\usepackage[table]{xcolor}
\usepackage{comment}
\usepackage{xcolor}
\usepackage{tikz}
\usetikzlibrary{arrows.meta, positioning}
\usepackage{booktabs} 
\usepackage{caption} 
\usepackage{subcaption} 
\usepackage{graphicx}
\usepackage{pgfplots}
\usepackage[all]{nowidow}
\usepackage[utf8]{inputenc}
\usepackage{tikz}
\usetikzlibrary{arrows.meta,decorations.pathreplacing}
\usetikzlibrary{backgrounds}
\usetikzlibrary{er,positioning,bayesnet}
\usepackage{multicol}
\usepackage{amsmath}
\usepackage{amssymb}
\usepackage{algpseudocode,algorithm,algorithmicx}
\usepackage{minted}
\usepackage[round,authoryear]{natbib}
\setcitestyle{authoryear,round}
\usepackage{hyperref}
\hypersetup{hidelinks}
\usepackage[inline]{enumitem} 
\definecolor{blue}{HTML}{1F77B4}
\definecolor{orange}{HTML}{FF7F0E}
\definecolor{green}{HTML}{2CA02C}

\pgfplotsset{compat=1.14}

\begin{document}
\title{Causal Regularization for Robust Prediction}
%
%
\author{Alessia Berarducci \inst{1} \and
Melania Lembo \inst{1} \and
Veronica Vinciotti \inst{2} \and Ernst C. Wit \inst{1}}
\authorrunning{F. Author et al.}
%
\institute{
USI, Università della Svizzera Italiana, Switzerland\\
\texttt{alessia.berarducci@usi.ch, melania.lembo@usi.ch, ernst.jan.camiel.wit@usi.ch}
\and
University of Trento, Italy\\
\texttt{veronica.vinciotti@unitn.it}
}
\maketitle      
%
\begin{abstract}

Prediction is an important use of a statistical model. However, predictive systems typically struggle when the environment, in which they were trained, changes. Robust prediction that can deal with distributional shifts has become an active research area. Existing approaches tend to make strong assumptions in order to achieve prediction guarantees. In this paper, we only require finite second moments in the training data to obtain sharp prediction guarantees. 


We generalize causal regularization to a framework that uses two environments and measures the differences between them through a generalized risk difference induced by changes in second-order moments of the covariates and response. 
By defining a sieve of out-of-sample distributions, we derive a worst-risk decomposition over increasing classes of environments and obtain a closed-form estimator that interpolates between OLS and the regularization limit solution. 
Causal interpretation is recovered under the inner-product invariance condition and when the conditional mean of the target given its causal parents remains invariant across environments.

Our results show that the proposed estimator can achieve prediction stability over a broad class of perturbed environments, even when trained on a limited set. The framework separates two goals: robustness as a stable prediction under weak assumptions and causality as a stronger property obtained with additional invariance structure.
\end{abstract}
\section{Introduction}

Prediction models are commonly trained under a limited set of conditions and subsequently applied in environments that may differ from those observed during training. Changes in populations, experimental conditions, or external interventions can alter the joint
distribution of the predictors and the response. Consequently, associations that provide accurate predictions in the training data may become unstable in new environments, leading to a substantial deterioration in predictive performance. 
This is an instance of a wider problem of out--of--distribution generalization that studies how models predict outcomes when the characteristics of the training data differ from the testing ones \citep{liu2023outofdistribution}. 
When data from multiple environments are available, their heterogeneity can provide information about the stability of a predictive relation. Differences across environments can help distinguish associations that depend on the observed data from relationships that remain stable when the data-generating system changes. This motivates the construction of predictors that retain good performance across environments rather than optimizing predictive accuracy in a single training distribution. This is the objective of this paper and what is referred to as robust prediction.

There is a connection between stability across environments and causality. Indeed, causal mechanisms are identified as processes that do not change unless we directly intervene on the outcome. Consequently, it is possible for causal relations to provide stable predictive information when the data distributions change. In that case, the patterns that stay invariant across different distributions are used to identify the underlying causal structure under suitable assumptions. This connection between invariance, causality and robustness provides a common perspective on causal inference and prediction under heterogeneous or perturbed data \citep{buhlmann2020invariance}. However, the resulting methods differ as their guarantees still depend on structural assumptions about the data-generating process and the interventions. Invariant causal prediction (ICP) \citep{peters2015causalinferenceusinginvariant} is an identification procedure: it attempts to find those predictors whose conditional relationship with the response remains invariant across different environments. Causal Dantzig \citep{rothenhausler2019causal}, similar to ICP, uses invariance via moment conditions, but relies on linear structural models and assumes that the interventions are sufficiently informative. These two methods have the same type of conditional identifiability issues when the observed environments do not provide enough variation or when their invariance assumptions are violated.
Anchor regression \citep{rothenhausler2021anchor} and causal regularization \citep{kania2025causalregularizationtradeoffinsample} represent an alternative approach to the two introduced above. Indeed, both methods take a more predictive and optimization-based perspective. However, anchor regression requires observing the anchor variables and is formulated within a linear model with additive shifts. Causal regularization requires an observational reference environment as well as a linear data-generating process. Both constraints will likely become restrictive when all observed environments are perturbed, when both predictors and the response are affected by interventions, or when the relationship among predictors and the response variable are non-linear.

In these settings, it remains meaningful to seek a set of predictors that is stable across environmental changes, even when a causal interpretation cannot be established.
Distributional robustness via invariant gradients (DRIG) \citep{shen2026causality} extends anchor regression by allowing additive random-noise interventions. Such interventions can induce shifts in both means and variances across environments. DRIG provides robustness against a data-dependent class of unseen perturbations, but relies on a linear structural
model and an observational reference environment. Likewise, \citet{christiansen2022causal} study robust prediction using a worst-case risk criterion, but require possible test distributions to be represented as interventions on a structural causal model rather than characterized through changes observed across environments.

In this paper, we extend causal regularization to robust linear prediction under more general forms of distributional heterogeneity. Although the proposed predictor is linear in its parameters, the joint data-generating mechanism of the predictors and the response need not be linear. 
The proposed criterion combines predictive accuracy in the two training environments with robustness to distributional changes. We show that it admits a worst-case risk interpretation over classes of unseen environments whose second-order moments are controlled by those observed during training. 
The resulting framework is predictive in its general form, while a causal interpretation is recovered under additional invariance conditions. We study separately the conditions under which the limiting parameter admits a causal interpretation. In particular, when the conditional mean of the response given its causal parents is invariant and the causal residual satisfies an inner-product invariance condition across environments, the causal parameter is a stationary solution of the original risk difference. If the difference between the covariate second-moment matrices is nonsingular, this solution is unique and coincides with the population limit of the regularization path.

Section~\ref{sec:robust-prediction} introduces the general robust prediction framework, develops its population and finite-sample formulations and describes the cross-validation procedure used to select the regularization parameter. Section~\ref{sec:causality_prediction_limit} establishes conditions under which the limiting robust parameter coincides with the causal parameter. Section~\ref{sec:application} illustrates the proposed method on a RPE1 single-cell perturbation data~\citep{replogle2022mapping} to study robust prediction and on the causal chamber light-tunnel experiment~\citep{gamella2025chamber} to investigate causal estimation.

\section{Robust prediction}
\label{sec:robust-prediction}

In this section, we define a completely general system $(X^e,Y^e)$, possibly affected by hidden confounding $H^e$, in  arbitrary observational or interventional data environments $e$. We extend the notion of causal regularization to this kind of possibly non-linear, non-additive and even non-causal systems. 
With two distinct data environments, we construct a regularization path that interpolates between the predictor minimizing the average risk across the observed environments and a limiting parameter determined by their distributional differences. The resulting predictor achieves sharp prediction bounds in certain unobserved out-of-distribution environments. 
In finite samples, the choice of the regularization parameter can be performed via cross-validation. 
Sections~\ref{subsec:2.1}--\ref{subsec:2.3} analyze the problem in a population setting, while Sections~\ref{subsec:finite-sample}--\ref{subsec:examples_robust} turn to the finite sample case.

\subsection{Data universe}
\label{subsec:2.1}

We consider a random vector $(X,Y)$ of interest, where $X\in \mathbb{R}^p$ are potential predictors and $Y\in\mathbb{R}$ the prediction target of interest. Rather than having a unique distribution, we assume that $(X,Y)$ may be observed according to various distributions, which we refer to as the \emph{data universe}. 

Formally, we first introduce the set of all $p+1$-dimensional distributions with finite second moments, 
\[
\mathcal{C}
=
\left\{
P \mid P \text{ is a probability distribution for } (X,Y) \text{ with finite second moment}
\right\}.
\]
Then, a \emph{data universe} $\mathcal{U}$ is any arbitrary subset of $\mathcal{C}$, i.e.,  \(\mathcal{U} \subset \mathcal{C}\) is the set of distributions of \((X,Y)\). This defines the \emph{data universe} as a collection of probability distributions according to which the system $(X,Y)$ may be observed. These individual probability measures $P^{X,Y}$ are sometimes referred to as \emph{environments}. 

It is useful to give some examples of possible data universes. 
\begin{itemize}
    \item \emph{Causal graphical model.} Consider a causal graphical model $(X,Y)\sim P$ according to a DAG $G$. Then hard do-interventions will result in a set of distributions $\mathcal{U}$ according to the causal factorization formula. 
    \item \emph{Unmeasured confounding.} Consider a distribution \(P^{X,Y,H}\) of the random vector \((X,Y,H)\), with values in $\mathbb{R}^p \times \mathbb{R} \times \mathbb{R}^q $, where \(H \in \mathbb{R}^q\) denotes hidden confounders affecting the joint behavior of \(X, Y\). Considering, for example, the set of hard, unobserved, interventions $H=h$, then by marginalizing over \(H\), we could obtain a set $\mathcal{U}$ of the marginal distributions \(P_h^{X,Y} \). These distributions do, in general, not satisfy any causal factorization properties. 
\end{itemize}
In section~\ref{subsec:examples_robust} we consider a number of even more specific examples. For now, it is important to realize that (1) each element of \(\mathcal{U}\) corresponds to one possible observable distribution or data environment, and (2) although this data universe may correspond to some causal system with interventions, this is not necessarily the case: the proposed framework does not require the observable distributions to be generated by an underlying causal system.

\subsection{Observed environments}

For the rest of the paper, we assume that the \emph{observed data} come from two environments \(e_1,e_2\), each characterized by different joint distributions \(P_{e_1},P_{e_2} \in \mathcal{U}\) of the random variables \((X^e,Y^e)\). Although it is, trivially, possible to associate with each distribution $P_e$ a DAG representing the conditional independence relations in a directed graphical model \citep{pearl2009causation}, there is no assumption in this section that the structure of the two DAGs associated with the observed environments are related to each other. Again, we stress that the data universe, in principle, does not have to have any causal interpretation.

The environment-specific distributions can be described in terms of their second-order moments. Accordingly, we associate each distribution \(P_e\) with the second moment matrix 
\[
M_e
=
\mathbb{E}
\left[
\begin{bmatrix}
X^e \\
Y^e
\end{bmatrix}
\begin{bmatrix}
X^e \\
Y^e
\end{bmatrix}^{t}
\right]
=
\begin{bmatrix}
G_e & Z_e \\
Z_e^t & S_e
\end{bmatrix},
\]
where \(G_e = \mathbb{E}[X^e X^{et}]\), \(Z_e = \mathbb{E}[X^eY^e]\) and \(S_e = \mathbb{E}[({Y^e})^2]\). For identifiability purposes, we will assume that either or both $G_{e_1}$ and $G_{e_2}$ are positive definite. This assumption is unrelated to the main focus of the paper. The notation \(G_e\), \(Z_e\) and \(S_e\) will be used throughout the following sections. We also define the sum and difference of these second order moment matrices, \[ M_{+} = M_{e_1}+M_{e_2}, \qquad M_{\Delta} = M_{e_1}-M_{e_2}. \] 
These aggregate matrices can be written as
\[
M_{+}
=
\begin{bmatrix}
G_{+} & Z_{+} \\
Z_{+}^{t} & S_{+}
\end{bmatrix},\quad
M_{\Delta}
=
\begin{bmatrix}
G_{\Delta} & Z_{\Delta} \\
Z_{\Delta}^{t} & S_{\Delta}
\end{bmatrix}
\]
where
$G_{+/\Delta}=G_{e_1} \pm G_{e_2},
Z_{+/\Delta}=Z_{e_1}\pm Z_{e_2}$, and
$S_{+/\Delta}=S_{e_1}\pm S_{e_2}$.

\subsection{Prediction in out-of-sample environments via robust regularization}
\label{subsec:2.3}
The aim is to predict \(Y\) from \(X\) using data from the two observed environments \(e_1\) and \(e_2\). Although we do not assume any linear relationship between \(X\) and \(Y\), we consider the set of all linear predictors. This restriction allows us to describe prediction performance through second-order moments and to compare environments by studying how risk functions vary between them. It is important to underline that the prediction model is linear in its coefficients, but, in principle, it need not be linear in the original covariates.

\begin{definition}[Within-environment risk, pooled risk and risk difference]
For an environment \(e\), we define the \emph{within-environment risk} of the linear predictor \(\widehat{Y}_e = \beta^t X^{e} \) as
\[
R_e(\beta)
=
\mathbb{E}\left[\left(Y^e - \beta^t X^{e}\right)^2\right]. 
\]
We define the \emph{pooled risk} $R_+$ and the \emph{risk difference} $R_\Delta$ as
\[
R_+(\beta)
=
R_{e_1}(\beta)
+
R_{e_2}(\beta),
\qquad
R_\Delta(\beta)
=
R_{e_1}(\beta)
-
R_{e_2}(\beta).
\]

 \end{definition}
The environmental risk $R_e$, the pooled risk $R_+$ and the risk difference $R_\Delta$ can all be written in matrix multiplication form relative to their associated second moment matrix, i.e.,
\[ R_{\star}(\beta) = [\beta, -1]~ M_{\star}~ [\beta, -1]^t.\]
The pooled risk \(R_+(\beta)\) aggregates the performance of the linear predictor over the observed environments and is minimized by the ordinary least squares $\beta_{OLS}$ in the pooled data. In contrast, the risk difference \(R_\Delta(\beta)\) quantifies how the prediction performance varies when moving from one environment to the other. In the special case of linear system with additive shifts and no intervention on the target, this risk difference has been shown to correspond to the objective function of the causal Dantzig estimator~\citep{rothenhausler2019causal}. This corresponds to the stationary points of the risk difference \(R_\Delta\). In our case, the risk difference does not necessarily have a minimum, but we introduce the \emph{generalized causal Dantzig} as its set of stationary points.   

\begin{definition}[Generalized causal Dantzig]
    The \emph{generalized causal Dantzig} population estimator is defined as the set of stationary points of the risk difference, i.e., 
\[
\mathcal{B}
=
\left\{
\beta \in \mathbb{R}^p
:
\nabla_\beta R_\Delta(\beta)=0
\right\}.
\]
\end{definition}
The following proposition establishes the relationship between the stationary points $\mathcal{B}$ and the second moment quantities.

\begin{proposition}
\label{prop:beta_cp}
The generalized causal Dantzig estimator corresponds to the set of  \(\beta \) that satisfy the linear system
\[
G_\Delta \beta = Z_\Delta.
\]
If \(G_\Delta\) is full rank, which requires that two data environments are sufficiently different in all covariate directions, then the system admits a unique solution 
\[
\beta^\star
=
G_\Delta^{-1}Z_\Delta.
\]

\end{proposition}

The proof is deferred to Appendix~\ref{appendix:betacpstatpoint}. If \(G_\Delta\) is singular, it is possible to show that we either have infinitely many solutions or $\mathcal{B}=\varnothing$. Under continuous and sufficient variation of the second-moment matrices across all covariate directions, singularity occurs only for particular combinations of the moment differences. However, it may arise naturally under structured interventions affecting only a lower-dimensional subset of the covariates. We retain the singular case in the subsequent analysis.
When \(\mathcal B=\varnothing\), the original covariate space may
be enlarged through a functional expansion of \(X\) \citep{wood2017generalized} ensuring that \(Z_\Delta \in \mathrm{Range} (G_\Delta)\).
To this end, let
$\mathcal F=\{f_j\}_{j\in\mathbb N}$ be a functional basis such that
$f_j:\mathcal X_1\times\cdots\times\mathcal X_p
\longrightarrow\mathbb R$
and
$f_j(X^{e_i})\in L_2,
i\in\{1,2\}.$
We assume that the environments \(e_i\), \(i\in\{1,2\}\), are such that there exists a common sequence of coefficients
\(\{\beta_j\}_{j\in\mathbb N}\) satisfying
$
\mathbb E\left[
Y^{e_i}
\mid
f_j(X^{e_i}),\ j\in\mathbb N
\right]
=
\sum_{j=1}^{\infty}
\beta_j f_j(X^{e_i})$ for $i\in\{1,2\},$
where the series converges in \(L_2\).
This allows for nonlinear prediction through, for example, spline expansions, where basis functions of \(X\) are used instead of the raw covariates. For the rest of the paper, we therefore will assume without loss of generality that $\mathcal{B}$ is not empty.

As both \(R_{e_1}(\beta)\) and \(R_{e_2}(\beta)\) are quadratic in \(\beta\), their difference \(R_\Delta(\beta)\) is also quadratic, albeit not necessarily convex. While in the special case of causal Dantzig the stationary point of \(R_\Delta (\beta)\) is attained in a minimum, for arbitrary environments \(e_1\) and \(e_2\) this is not necessarily the case. We define a convexified version of the risk difference, which restores this connection and provides the theoretical guarantees required by the robust formulation.
To this end, let the spectral decomposition of \(G_\Delta\) be $
G_\Delta = V\Lambda V^t$,
and denote 
\[
G_\Delta^{+}=V|\Lambda|V^t,
\]
as the positive semidefinite matrix  obtained by replacing the eigenvalues in \(\Lambda\) with their absolute values.

\begin{definition}[Convexified risk difference]
    The convexified risk difference centered at \(\beta^\star\in \mathcal{B}\) is defined as
\[
R_{\Delta}^{+}(\beta)
=
(\beta-\beta^\star)^t
G_\Delta^{+}
(\beta-\beta^\star).\]
\end{definition} 
By defining \(Z_\Delta^+ = G_\Delta^+\beta^\star\) and \(S_\Delta^+ = {\beta^\star}^t G_\Delta^+\beta^\star\), the same quantity can be expressed as
$R_{\Delta}^{+}(\beta)
=
\beta^t G_\Delta^{+}\beta
-
2\beta^t Z_\Delta^{+}
+
S_\Delta^{+}$. Introducing the positive semi-definite second moment difference matrix 
\[
M_{\Delta}^{+}
=
\begin{bmatrix}
G_\Delta^{+} & Z_{\Delta}^{+} \\[1ex]
Z_{\Delta}^{+ t} & S_{\Delta}^{+}
\end{bmatrix},
\]
we can write the convexified risk difference as 
\[
R_{\Delta}^{+}(\beta)
= [\beta, -1]~ M_\Delta^+ ~[\beta, -1]^t
\]
The convexified risk difference captures the perturbations induced jointly by the two environments while ensuring positive semi-definiteness. As the next proposition shows, the convexified risk does not depend on which element $\beta^\star$ has been selected from $\mathcal{B}$. 

\begin{proposition}
\label{prop:min_convex}
The convexified risk $R_\Delta^+$ does not depend on the element $\beta^\star\in \mathcal{B}$ selected in its definition. 
\end{proposition}

%


The proof is provided in Appendix~\ref{appendix:kernel}, but fundamentally relies on the fact that \(G_\Delta\) and \(G_\Delta^+\) have the same kernel. Since $R_\Delta^+(\beta)\geq 0$, its minimum is attained precisely when
$\beta-\beta^\star\in\ker(G_\Delta^+)
=
\ker(G_\Delta)$,
which is equivalent to
$G_\Delta\beta=Z_\Delta$. Therefore, we obtain another way to characterize $\mathcal{B}$ via the convexified risk difference,
\[
\mathcal B = \arg\min_{\beta\in\mathbb R^p}
R_\Delta^+(\beta).
\]

While ideally we would select \(\beta\) to minimize the predictive risk over the full data universe \(\mathcal{U}\), we begin by limiting our focus to a smaller class of settings that exist outside of our current data samples. These environments represent plausible perturbations of the two observed training environments \(e_1\) and \(e_2\) in ways that are likely to occur. 

\begin{definition}[Out-of-distribution environments] The $\gamma$-perturbed out-of-distribution distribution set,
\begin{equation}
\mathcal{C}_{\gamma}
=
\left\{
\, P_{e}^{X,Y}\in\mathcal C
\;\middle|\;
M_{e}
\preceq
\frac{1}{2}M_+
+
\frac{\gamma}{2}M_\Delta^+
\right\},
\label{eq:cgamma}
\end{equation}
is defined as the set of distributions of $(X,Y)$ whose second moment matrix $M_e$ are bounded, in the Loewner sense, by a linear combination of the average in-sample second moment matrix and the $\gamma$-inflated positive semi-definite second moment difference matrix for any $\gamma\geq 0$. 
\end{definition}

By construction,  \(\mathcal{C}_{\gamma}\) is a subset of the class of distributions under consideration, i.e., \(\mathcal{C}_{\gamma} \subset \mathcal{C}\). The parameter \(\gamma \) controls the size of the admissible perturbation class. When \(\gamma\) is small, we consider environments close to the pooled training data. As \(\gamma\) increases, we allow stronger deviations in the directions described by the differences
between \(e_1\) and \(e_2\), through the matrix \(M_{\Delta}^{+}\). 
If the perturbation model is sufficiently rich to describe all significant interventions on the system, then the limiting robust class may coincide with the full admissible class of distributions \(\mathcal{C}\). The next result allows us to explicitly write a tight bound to the worst possible risk within the out-of-distribution set $\mathcal{C}_\gamma$.

\begin{proposition}[Worst-risk decomposition]
\label{prop:worst_case}
For every \(\beta \in \mathbb{R}^p\), the worst-case risk over \( \mathcal{C}_{\gamma} \) satisfies
\begin{equation}
\sup_{P_{e}\in \mathcal C_\gamma}
R_{e}(\beta)
=
\frac{1}{2}R_+(\beta)
+
\frac{\gamma}{2}R_\Delta^+(\beta).
\end{equation}

\end{proposition}
The full proof is provided in Appendix~\ref{appendix:worst_case}. The result follows from the definition of \(\mathcal{C}_{\gamma}\) and the quadratic representation of the risk. In particular, the upper bound on the second-order moment matrix induces an upper bound on the corresponding quadratic risk. Equality is achieved by constructing an explicit distribution whose second-order moments matches the upper bound. 

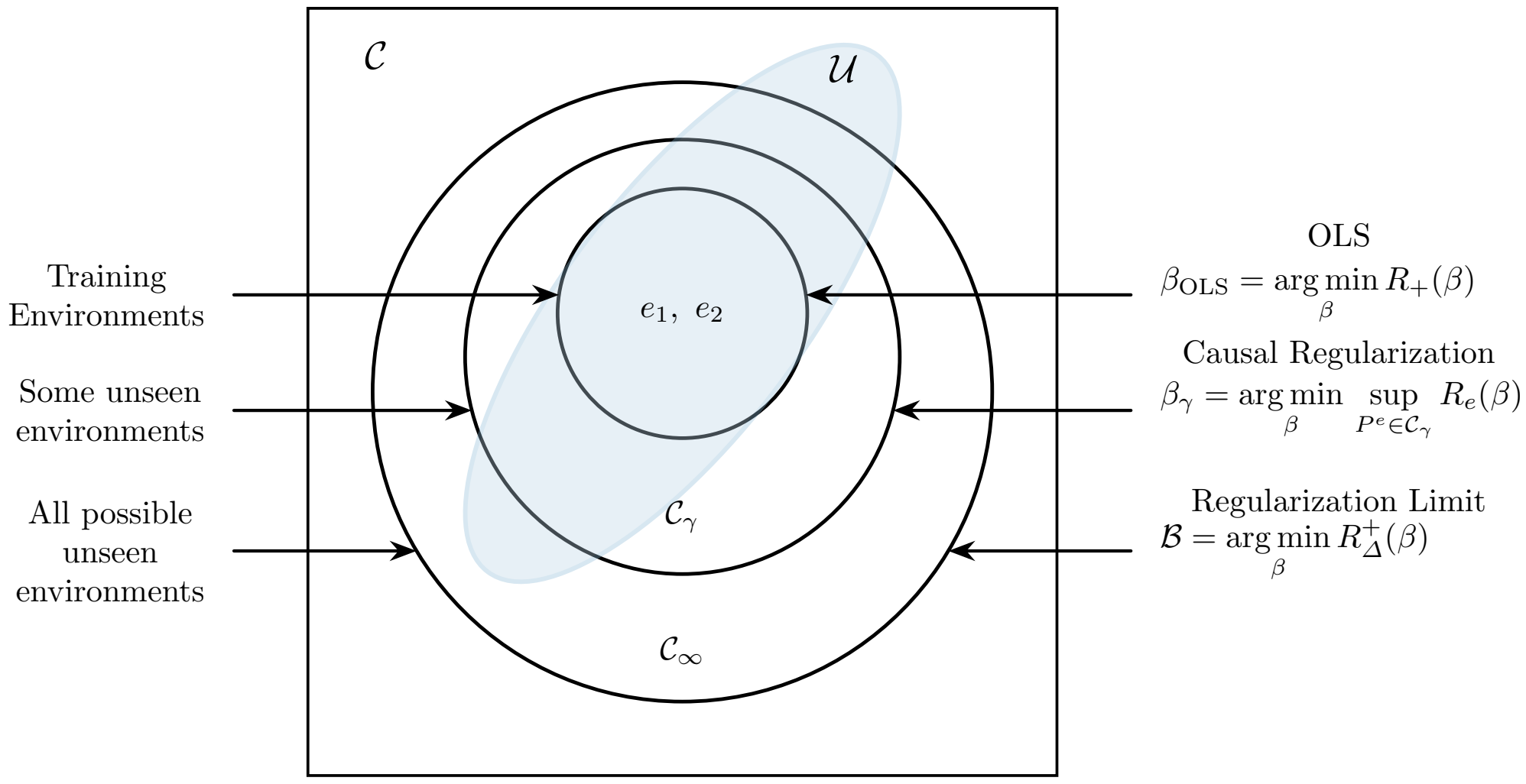
\begin{figure}[t]
    \centering
\begin{tikzpicture}[
  scale=0.8,
  transform shape,
  >=Stealth,
  line cap=round,
  every node/.style={font=\normalsize},
  arrow/.style={->,line width=0.8pt}
]
  \def\yinner{0.85}
  \def\ymiddle{0.38}
  \def\youter{0.00}
  \coordinate (I) at (0,\yinner);
  \coordinate (M) at (0,\ymiddle);
  \coordinate (O) at (0,\youter);
  \def\rinner{1.35}
  \def\rmiddle{2.35}
  \def\router{3.35}

  \draw[line width=0.7pt] (-4.05,-4.15) rectangle (4.05,4.15);
  \node[anchor=north west,font=\large] at (-3.55,3.90) {$\mathcal C$};
  \draw[line width=0.9pt] (O) circle[radius=\router];
  \draw[line width=0.9pt] (M) circle[radius=\rmiddle];
  \draw[line width=0.9pt] (I) circle[radius=\rinner];
\filldraw[
  fill=blue!30,
  draw=blue!50,
  line width=1.1pt,
  opacity=0.35
] (0,0.85) ellipse [x radius=1.3, y radius=3.5, rotate=-37];

 \node[font=\large] at (1.75,3.5) {$\mathcal U$};
  \node at (0,0.85) {$e_1,\ e_2$};
  \node at (0,-1.35) {$\mathcal C_\gamma$};
  \node at (0,-2.80) {$\mathcal C_\infty$};

  \def\ytop{1.05}
  \def\ymid{-0.2}
  \def\ybot{-1.72}
  \pgfmathsetmacro{\xtop}{sqrt(\rinner*\rinner-(\ytop-\yinner)^2)}
  \pgfmathsetmacro{\xmid}{sqrt(\rmiddle*\rmiddle-(\ymid-\ymiddle)^2)}
  \pgfmathsetmacro{\xbot}{sqrt(\router*\router-(\ybot-\youter)^2)}

  \node[align=center,anchor=east] at (-5.05,\ytop)
    {Training\\Environments};
  \node[align=center,anchor=east] at (-5.05,\ymid)
    {Some unseen\\environments};
  \node[align=center,anchor=east] at (-5.05,\ybot)
    {All possible\\unseen\\environments};

  \draw[arrow] (-4.85,\ytop) -- (-\xtop,\ytop);
  \draw[arrow] (-4.85,\ymid) -- (-\xmid,\ymid);
  \draw[arrow] (-4.85,\ybot) -- (-\xbot,\ybot);
  \draw[arrow] (4.85,\ytop) -- (\xtop,\ytop);
  \draw[arrow] (4.85,\ymid) -- (\xmid,\ymid);
  \draw[arrow] (4.85,\ybot) -- (\xbot,\ybot);

  \node[anchor=west] at (5.05,\ytop)
    {$\displaystyle \beta_{\mathrm{OLS}}
      =\mathop{\arg\min}_{\beta}R_{+}(\beta)$};
  \node[anchor=west] at (5.05,\ymid)
    {$\displaystyle \beta_{\gamma}
      =\mathop{\arg\min}_{\beta}\,\sup_{P^e\in\mathcal C_{\gamma}}R_e(\beta)$};
  \node[anchor=west] at (5.05,\ybot)
    {$\displaystyle \mathcal{B}
      =\mathop{\arg\min}_{\beta}R_{\Delta}^{+}(\beta)$};

  \node at (7.10,1.7) {OLS};
  \node at (7.10,0.4) {Causal Regularization};
    \node at (7.10,-1.2) {Regularization Limit};
\end{tikzpicture}

\caption{The robust class \(\mathcal{C}_{\gamma}\) contains plausible unseen shifts. The shadowed region \(\mathcal{U}\) denotes the data universe. The environments \(e_1\) and \(e_2\) are the observed training environments, while the outer region represents all possible future environments.}
  \label{fig:robust_classes}
\end{figure}

Figure \ref{fig:robust_classes} summarizes the nested structures of interest, starting from the observed environments at the center to the full admissible class at the outer region. The parameters \(\gamma\) controls the size of \(C_\gamma\), which expands as \(\gamma\) increases.  
Following Proposition~\ref{prop:worst_case}, the associated robust estimator is obtained by minimizing this worst-case risk.

\begin{definition}[$\gamma$-robust estimator]
For a given robustness $\gamma\geq 0$, the robust estimator $\beta_\gamma$ is defined as any minimizer of the worst-case risk over the global robust class $\mathcal{C}_{\gamma}$, namely,
\begin{equation}
\beta_\gamma
= 
\arg\min_{\beta \in\mathbb{R}^p} 
\left\{
\frac{1}{2}R_{+}(\beta)
+
\frac{\gamma}{2}R_{\Delta}^{+}(\beta)
\right\}.
\end{equation}
\end{definition}
Consequently, \(\beta_\gamma\) is the linear parameter minimizing the worst-case prediction risk over the admissible class \(\mathcal{C}_{\gamma}\). The regularization parameter \(\gamma\) controls the trade-off between in-sample predictive accuracy and robustness to distributional perturbations. When \(\gamma=0\), the estimator reduces to the pooled ordinary least squares solution, \(\beta_{OLS}\). In the limit, the solution corresponds to the generalized causal Dantzig estimator, $\lim_{\gamma\to\infty}\beta_\gamma=\mathcal{B}$. For $0<\gamma<\infty$, the estimator \(\beta_\gamma\) balances fitting the observed environments and possible future distribution in $\mathcal{U} \cap \mathcal{C}_\gamma$. For each $\gamma\geq 0$, the full rank of at least one of the Grammians $G_{e_1}$ and $G_{e_2}$, guarantees that there is an explicit and unique solution, given by
\[ \beta_\gamma = G_\gamma^{-1} Z_\gamma,\]
where $G_\gamma= \frac{1}{2}G_{+}
+
\frac{\gamma}{2}G_{\Delta}^{+}
$ and $Z_\gamma= \frac{1}{2}Z_{+}
+
\frac{\gamma}{2}Z_{\Delta}^{+}
$. The proof is given in Appendix \ref{appendix:beta_gamma}.



\subsection{Finite-sample robust prediction}
\label{subsec:finite-sample}

The results described in the previous section depend on the population moments of the joint distributions of \(X\) and \(Y\). In practice, these moment matrices are not directly available and need to be estimated from finite samples. In this section, we define the finite sample plug-in estimator $\hat\beta_\gamma$ and show that it is consistent.

For \(e\in\{e_1,e_2\}\), let \(\mathcal{D}_e=\{(X_i^e,Y_i^e)\}_{i=1}^{n_e}\)  denote the finite samples collected in two observed data environments. For $e\in\{e_1,e_2\}$, we denote by \(\mathbf{X}_e\in\mathbb{R}^{n_e\times p}\) and \(\mathbf{y}_e\in\mathbb{R}^{n_e}\) the two covariate matrices and response vectors. We replace the population moments by their empirical counterparts, computed by averaging over the observations within each environment. For \(e\in\{e_1,e_2\}\), we define
\[
\mathbb{Z}_e
=
\frac{1}{n_e}\mathbf{X}^{t}_e\mathbf{y}_e, \quad 
\mathbb{G}_e
=
\frac{1}{n_e}\mathbf{X}^{t}_e\mathbf{X}_e.
\]
The empirical sum and difference moments are defined as
$
\mathbb{Z}_{+}
=
\mathbb{Z}_{e_1}+\mathbb{Z}_{e_2},
$
$\mathbb{Z}_{\Delta}
=
\mathbb{Z}_{e_1}-\mathbb{Z}_{e_2},
$ 
and
$
\mathbb{G}_{+}
=
\mathbb{G}_{e_1}+\mathbb{G}_{e_2},
$
$\mathbb{G}_{\Delta}
=
\mathbb{G}_{e_1}-\mathbb{G}_{e_2}.$ 
For both environments \(e\in\{e_1,e_2\}\), the empirical prediction risk is defined as
\[
\mathbb{R}_e(\beta)
=
\frac{1}{n_e}
\sum_{i=1}^{n_e}
\left(
Y_i^e-\beta^t X_i^e
\right)^2.\]
The empirical pooled risk and empirical risk difference are defined as $\mathbb{R}_{+}(\beta)
=
\mathbb{R}_{e_1}(\beta)
+
\mathbb{R}_{e_2}(\beta)$ and $\mathbb{R}_{\Delta}(\beta)
=
\mathbb{R}_{e_1}(\beta)
-
\mathbb{R}_{e_2}(\beta)$, respectively. The empirical generalized causal Dantzig estimator is defined as the set of stationary points of the risk difference, which corresponds to
\[
\widehat{\mathcal B}
=
\left\{
\beta\in\mathbb R^p
\;\middle|\;
\mathbb{G}_{\Delta}\beta
=
\mathbb{Z}_{\Delta}
\right\}.
\]
When  $\mathbb{G}_\Delta$ has several zero eigenvalues, the equation defining $\widehat{\mathcal B}$ may admit multiple solutions. Following \cite{kania2025causalregularizationtradeoffinsample}, we identify -- purely for practical reasons -- the minimum Euclidean-norm solution, i.e., $\operatorname*{arg\,min}_{\beta\in\widehat{\mathcal B}}
\|\beta\|_2$, which is uniquely given as $\widehat\beta^\star= \mathbb{G}_\Delta^\dagger\mathbb{Z}_\Delta$, where $\mathbb{G}_\Delta^\dagger$ denotes the Moore-Penrose pseudoinverse. Given the eigen decomposition $\mathbb{G}_{\Delta}^{+}
=
\mathbb{V}
\mathbb{L}
\mathbb{V}^{t}$, we define the positive definite matrix
\[
\mathbb{G}_{\Delta}^{+}
=
\mathbb{V}
\left|\mathbb{L}\right|
\mathbb{V}^{t},
\]
where $\left|\mathbb{L}\right|$ is a diagonal matrix consisting of the absolute values of the eigenvalues in $\mathbb{L}$. Using $\widehat{\beta}^\star$ and $\mathbb{G}_\Delta^+$, we define the \emph{empirical convexified risk} as 
\[
\mathbb{R}_{\Delta}^{+}(\beta)
=
(\beta-\widehat\beta^\star)^t
\mathbb{G}_{\Delta}^{+}
(\beta-\widehat \beta^\star).
\]
This then allows us to define the finite-sample estimator set as
\[
\widehat{\mathcal{B}}_{\gamma}
=
\arg \min_{\beta\in\mathbb R^p} 
\frac{1}{2}\mathbb{R}_{+}(\beta)
+
\frac{\gamma}{2}
\mathbb{R}_{\Delta}^{+}(\beta).
\]
Just as in the population scenario, the objective is quadratic and its minimum is a solution of $\mathbb{G}_\gamma \beta = \mathbb{Z}_\gamma$, where \(\mathbb{G}_{\gamma} =\frac{1}{2} \mathbb{G}_{+} + \frac{\gamma}{2}\mathbb{G}_{\Delta}^{+}, \) and \(\mathbb{Z}_{\gamma} = \frac{1}{2}\mathbb{Z}_{+} + \frac{\gamma}{2}\mathbb{G}_{\Delta}^{+}\widehat{\beta}^\star\). For a finite sample, there is no guarantee that $\widehat{\mathcal{B}}_\gamma$ is unique. For this reason, we select the parameter vector with the minimum Euclidean norm using the Moore-Penrose inverse $\mathbb{G}_\gamma^\dagger$ to prove consistency. 


\begin{proposition}[Consistency of the finite-sample estimator]
For a fixed \(\gamma\geq0\), the minimum Euclidean norm estimator,
\[ \widehat\beta_\gamma = \mathbb{G}_\gamma^\dagger \mathbb{Z}_\gamma, \]
is consistent, i.e.,
\[
\widehat\beta_\gamma
\xrightarrow{P_{e_1},P_{e_2}}
\beta_\gamma
\]
as \(n_{e_1},n_{e_2}\to\infty\).
\end{proposition}

The proof is given in Appendix~\ref{appendix:consistency}. In practice, it may be useful to have some idea about how far off the out-of-distribution estimator is from its population. The next result derives the variance of the regularized estimator conditional on the in-sample covariates $\mathbf X=(\mathbf{X}_{e_1},\mathbf X_{e_2})$.

\begin{proposition}[Conditional Variance of the finite-sample estimator]
\label{prop:conditional-variance}
For fixed $\gamma\geq0$, the estimator
$\widehat\beta_\gamma=\mathbb G_\gamma^\dagger\mathbb Z_\gamma$
has conditional covariance
\[
\operatorname{Var}
(\widehat\beta_\gamma\mid \mathbf X)
= \frac{1}{4}
\mathbb G_\gamma^\dagger
\left[(I+\gamma \mathbb G_\Delta^+ \mathbb G_\Delta^\dagger) \Sigma_+(I+\gamma \mathbb G_\Delta^\dagger \mathbb G_\Delta^+) +2\mathbb G_\Delta^+ \mathbb G_\Delta^\dagger\Sigma_\Delta - \gamma(\mathbb G_\Delta^+ \mathbb G_\Delta^\dagger\Sigma_+ + \Sigma_+ \mathbb G_\Delta^\dagger \mathbb G_\Delta^+) \right]
\mathbb G_\gamma^\dagger.
\]
where we define $\Sigma_+ = \mathbf X_{e_1}^t V(\mathbf Y_{e_1} \mid \mathbf X_{e_1}) \mathbf X_{e_1} + \mathbf X_{e_2}^t V(\mathbf Y_{e_2} \mid \mathbf X_{e_2}) \mathbf X_{e_2}$ and $\Sigma_\Delta$ its corresponding difference.
\end{proposition}
For the conditional variance of the regularized estimator, we need 
$\operatorname{Var}(\mathbb Z_\gamma\mid \mathbf X)$. The derivation of this quantity is provided in Appendix~\ref{appendix:conditional-variance}. In practice, we are required to estimate the conditional variance of $\mathbf Y$ given $\mathbf X$. One option is to replace the variance matrix $V(\mathbf Y_{e} \mid \mathbf X_{e})$ by a diagonal matrix, whose diagonal consists of the squared values of $\mathbf Y_e - \mathbf X_e \hat \beta_\gamma$.

\subsection{Cross-Validation}
\label{sec:cross_validation}
 We employ cross-validation as a principled strategy for model selection. The parameter \(\gamma\) regulates the trade-off between pooled predictive accuracy and stability across shifted environments. When \(\gamma=0\), the estimator coincides with the ordinary least-squares solution
\(\widehat\beta_{{OLS}}\), which minimizes the empirical pooled risk without explicitly accounting for heterogeneity between the environments. As \(\gamma\) increases, greater weight is assigned to the convexified risk difference and the estimator increasingly approaches the stationary set \(\widehat{\mathcal B}\).

The observations from the two training environments are partitioned into a number of approximately equal-sized folds. For each fold and each candidate value of \(\gamma\), we estimate the model on the training folds and evaluate the convexified risk difference using the remaining folds. The fold-specific values are averaged and the selected value \(\gamma_{CV}\) is the one that achieves the smallest average held-out convexified risk difference. The final estimator is subsequently recomputed on the complete samples using \(\gamma_{CV}\). We minimize \(R^{+}_{\Delta}\) because it measures the
distance from its stationary point \(\beta^\star\). Smaller values of the convexified risk difference imply
greater stability when inputs vary. As a consequence, the resulting predictor remains robust over a larger class of environments.

The selected value of \(\gamma\) may be affected by how precisely the stationary solution can be estimated. When the two environments are similar, their moment differences are weak and the estimate of \(\widehat\beta_{RL}\) may be sensitive to sampling variation. In this case, cross-validation may favor an intermediate value of \(\gamma\). When the environments differ more clearly across the relevant covariate directions, \(\widehat\beta_{RL}\) can be estimated more reliably and larger values of \(\gamma\) may be favoured.


\subsection{Illustrative examples}
\label{subsec:examples_robust}
We consider two illustrative data-generating systems to examine the behavior of the proposed method. The structural parameters of these systems are assumed to be known only for the purpose of defining the examples. The objective is not to estimate the causal parameters. Rather, the examples are used to study the estimation of the robust parameter $\widehat{\beta}_{\gamma}$ from finite samples under different interventions or distributional shifts.

\subsubsection{Robust prediction does not necessarily imply causality.}

We consider two environments with an unobserved variable \(H\) that confounds the relationship between \(X_1\) and \(Y\). The intervention in the second environment shifts \(X_1\), \(X_2\) and \(Y\), while the hidden confounding structure remains unchanged. Figure~\ref{fig:causal_models_1} shows the causal graphs and the corresponding structural equations for the two environments. In each environment \(e\), the data are generated according to
\[
\begin{pmatrix}
Y^e\\
X_1^e\\
X_2^e\\
H^e
\end{pmatrix}
=
\left(I-B_{YXH}\right)^{-1}
\left[
\begin{pmatrix}
\varepsilon_Y^e\\
\varepsilon_1^e\\
\varepsilon_2^e\\
\varepsilon_H^e
\end{pmatrix}
+
\begin{pmatrix}
A_Y^e\\
A_1^e\\
A_2^e\\
A_H^e
\end{pmatrix}
\right],
\]
where
where the matrix \(B_{YXH}\) specifies the structural causal model. 
\[
B_{YXH}
=
\begin{pmatrix}
 0 & 1 &  0 & 1\\
 0 & 0 &  0 & 1\\
-1 & 0 &  0 & 0\\
 0 & 0 &  0 & 0
\end{pmatrix}.
\]
\begin{figure}[t]
\centering

\tikzset{
  causalnode/.style={
    font=\large,
    inner sep=2pt
  },
  observational/.style={
    -{Latex},
    thick,
    black
  },
  intervention/.style={
    -{Latex},
    thick,
    dashed,
    red!80!black
  }
}

\begin{minipage}[t]{0.48\textwidth}
\vspace{0pt}
\centering

\textbf{Environment $(e_1)$}\\[6pt]

\begin{tikzpicture}
  \node[causalnode] (H1)  at (0,1.4) {$H$};
  \node[causalnode] (X11) at (-0.9,0) {$X_1$};
  \node[causalnode] (Y1)  at (0.9,0)  {$Y$};
  \node[causalnode] (X21) at (2.5,0)  {$X_2$};

  \draw[observational] (H1)  -- (X11);
  \draw[observational] (H1)  -- (Y1);
  \draw[observational] (X11) -- (Y1);
  \draw[observational] (Y1)  -- (X21);

  \node[
    font=\normalsize,
    text=gray!70!black,
    align=center
  ] at (0.8,-1.2) {
    $\varnothing$\\[-2pt]
    \scriptsize No interventional shifts
  };

  \node[causalnode, opacity=0] at (-0.9,-1.2) {$A_1$};
  \node[causalnode, opacity=0] at (0.9,-1.2)  {$A_Y$};
  \node[causalnode, opacity=0] at (2.5,-1.2)  {$A_2$};
\end{tikzpicture}

\[
\begin{gathered}
H^{e_1}=2+\varepsilon_H,\\
X_1^{e_1}=H^{e_1}+\varepsilon_1^{e_1},\\
Y^{e_1}=X_1^{e_1}+H^{e_1}+\varepsilon_Y^{e_1},\\
X_2^{e_1}=-Y^{e_1}+\varepsilon_2^{e_1}.
\end{gathered}
\]

\end{minipage}
\hfill
\begin{minipage}[t]{0.48\textwidth}
\vspace{0pt}
\centering

\textbf{Environment $(e_2)$}\\[6pt]

\begin{tikzpicture}
  \node[causalnode] (H2)  at (0,1.4) {$H$};
  \node[causalnode] (X12) at (-0.9,0) {$X_1$};
  \node[causalnode] (Y2)  at (0.9,0)  {$Y$};
  \node[causalnode] (X22) at (2.5,0)  {$X_2$};

  \node[causalnode] (A1) at (-0.9,-1.2) {$A_1$};
  \node[causalnode] (AY) at (0.9,-1.2)  {$A_Y$};
  \node[causalnode] (A2) at (2.5,-1.2)  {$A_2$};

  \draw[observational] (H2)  -- (X12);
  \draw[observational] (H2)  -- (Y2);
  \draw[observational] (X12) -- (Y2);
  \draw[observational] (Y2)  -- (X22);

  \draw[intervention] (A1) -- (X12);
  \draw[intervention] (AY) -- (Y2);
  \draw[intervention] (A2) -- (X22);
\end{tikzpicture}

\[
\begin{gathered}
H^{e_2}=2+\varepsilon_H,\\
X_1^{e_2}=H^{e_2}+\varepsilon_1^{e_2}+A_1^{e_2},\\
Y^{e_2}=X_1^{e_2}+H^{e_2}
+\varepsilon_Y^{e_2}+A_Y^{e_2},\\
X_2^{e_2}=-Y^{e_2}+\varepsilon_2^{e_2}+A_2^{e_2}.
\end{gathered}
\]

\end{minipage}

\caption{Causal graphs and corresponding structural models for environments
$(e_1)$ and $(e_2)$. Black arrows represent the invariant relationships,
whereas red dashed arrows represent the interventional shifts in $(e_2)$.
All noise variables are mutually independent and distributed as
$\mathcal{N}(0,1)$. The interventions satisfy
$A_1^{e_2}\sim\mathcal{N}(2,1)$,
$A_2^{e_2}\sim\mathcal{N}(1,1)$, and
$A_Y^{e_2}\sim\mathcal{N}(3,4)$.}
\label{fig:causal_models_1}

\end{figure}
This example is adapted from the original causal regularization setting \citep{kania2025causalregularizationtradeoffinsample}, in which \(e_1\) is taken as the observational environment, so that \(A^{e_1}=0\), while non-zero shifts \(A^e=(A_1^e,A_2^e,A_Y^e)^t\) are introduced in \(e_2\). We retain this construction for the numerical illustration, although the current framework does not require either environment to be observational. The true causal coefficient (CP) for predicting \(Y\) from \(X=(X_1,X_2)^t\) is
\[
\beta_{{CP}}
=
\begin{bmatrix}
1\\
0
\end{bmatrix}.
\]
In particular, \(X_2\) is not a cause of \(Y\), even though it can be highly predictive because it is generated as a child of \(Y\).

\begin{figure}[t]
    \centering
    \begin{minipage}[t]{0.49\textwidth}
        \centering
        \includegraphics[width=\linewidth]{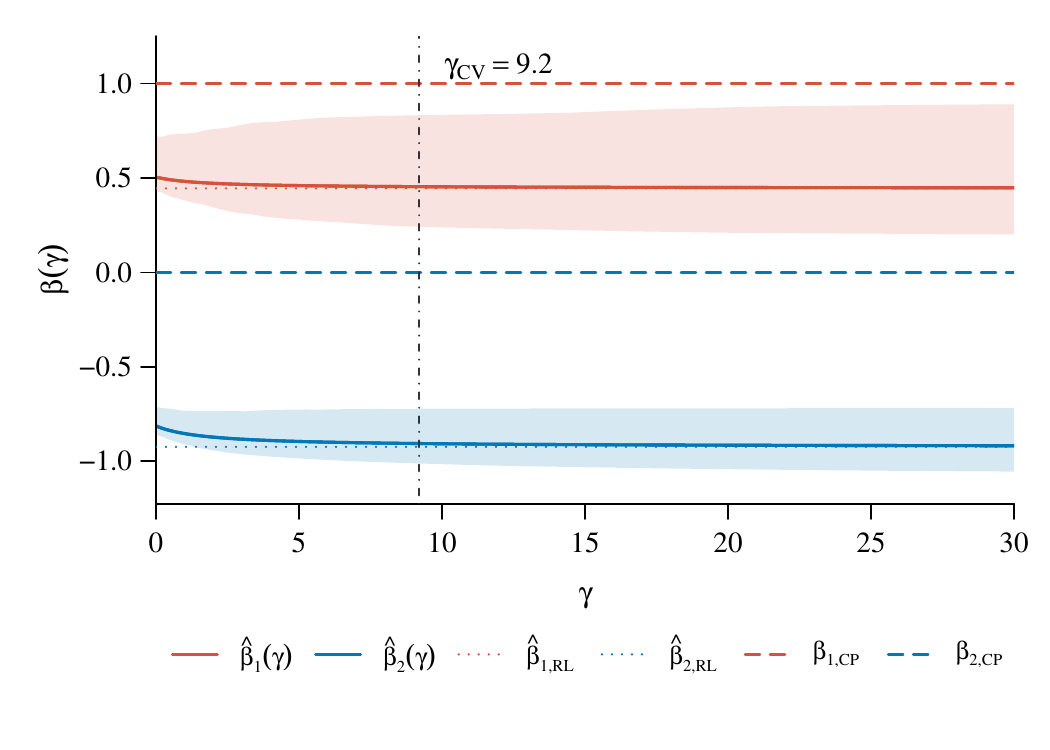}
        \caption*{\textbf{(a)} Regularization path.}
    \end{minipage}
    \hfill
    \begin{minipage}[t]{0.49\textwidth}
        \centering
        \includegraphics[width=\linewidth]{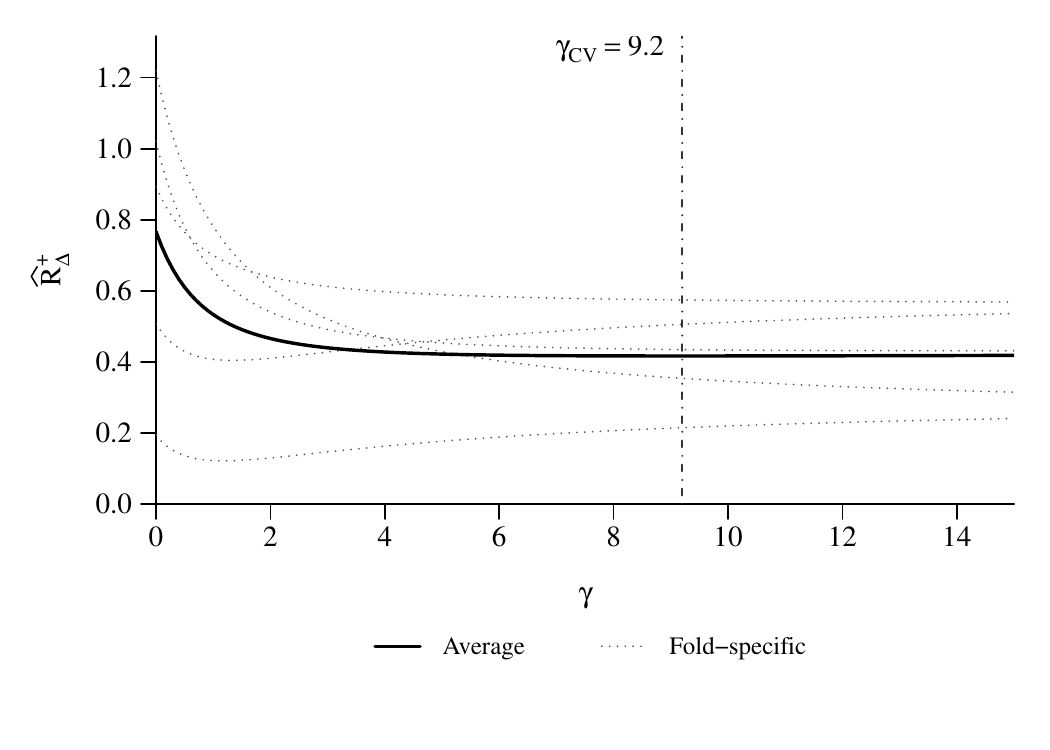}
        \caption*{\textbf{(b)} Cross-validation criterion}
    \end{minipage}
\caption{\textbf{(a)} The solid curves represent the components of \(\widehat{\beta}_{\gamma}\) and the shaded regions show the corresponding \(95\%\) pointwise confidence bands. The dotted horizontal lines denote the robust reference parameter \(\widehat{\beta}_{RL}\), while the dashed horizontal lines denote the true causal coefficient \(\beta=(1,0)^t\). The vertical dot-dashed line marks the selected value \(\gamma_{CV}=9.2\). \textbf{(b)} Fold-specific and average values of \(\mathbb{R}_{\Delta}^{+}(\widehat{\beta}_{\gamma})\) as a function of \(\gamma\). The vertical dot-dashed line marks the minimizing value
\(\gamma_{CV}=9.2\).}
   \label{fig:exampleB_results}
\end{figure}

Figure~\ref{fig:exampleB_results} shows how the two estimated coefficients change as the penalty parameter varies (left panel). The solid lines represent the estimated coefficients, while the shaded regions show their pointwise confidence bands. The robust reference parameter and the true causal coefficient are also shown. The right panel reports the cross-validation criterion for each fold and the average across folds.

The results highlight the distinction between robustness and causality. The  regularized estimator \(\widehat{\beta}_{\gamma}\) is not constructed to recover the causal coefficient under arbitrary shifts. Rather, it minimizes a worst-case prediction criterion that balances pooled in-sample fit and stability with respect to the observed distributional change. In particular, the second environment includes a direct additive shift on \(Y\), which introduces variation in the response that is not generated by its observed causal parent \(X_1\). Since \(X_2\) is a child of \(Y\), this shift is also propagated to \(X_2\), making it potentially useful for robust prediction despite having no causal effect on \(Y\).

Consequently, the regularization path moves towards the regularization limit parameter \(\widehat \beta_{RL}\), rather than towards the true causal value \(\beta_{CP}=(1,0)\). The fact that \(\widehat{\beta}_{RL}=(0.44,-0.92)^t\) differs from \(\beta_{CP}\) shows that, in this setting, the robust prediction target and the causal target are distinct.
The selected value \(\gamma_{CV}=9.2\) corresponds to the regularization parameter that minimizes the convexified risk-difference criterion \(\mathbb{R}^{+}_{\Delta}(\widehat{\beta}_{\gamma})\). The corresponding
estimator is the one that is most stable across the two environments according to this criterion.

\subsubsection{Robust prediction under a non-causal system.}
We next consider an even more general setting in which the two environments need not arise from a common structural model or share the same causal graph. Both the covariate distributions and the mechanisms relating \(X_1\), \(Y\) and \(X_2\) may change across environments.
In particular, in environment \(e_1\), \(X_1\) determines \(Y\), which in turn affects \(X_2\). In environment \(e_2\), both \(X_1\) and \(X_2\) determine \(Y\). Moreover, the response mechanism is nonlinear and differs between the two environments. The corresponding data generating processes, together with their DAGs, are reported in Figure ~\ref{fig:exampleB_models}.

\begin{figure}[t]
    \centering
    \begin{minipage}[t]{0.48\textwidth}
        \centering
        \includegraphics[width=\linewidth]{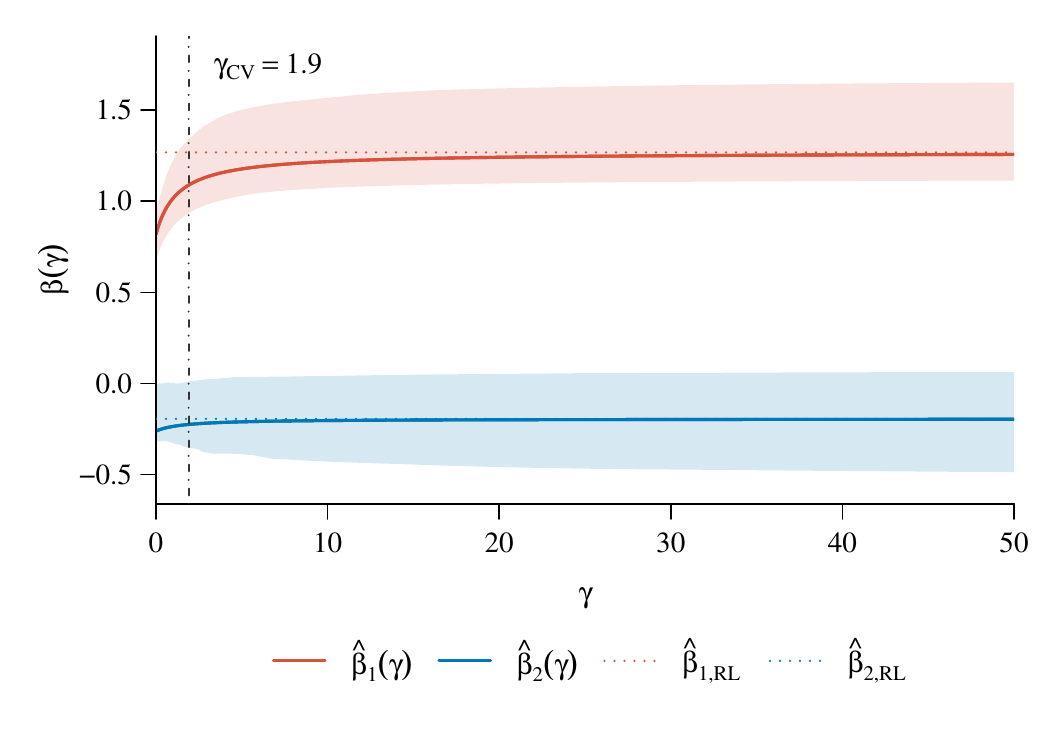}
        \caption*{\textbf{(a)} Regularization path}
    \end{minipage}
    \hfill
    \begin{minipage}[t]{0.48\textwidth}
        \centering
        \includegraphics[width=\linewidth]{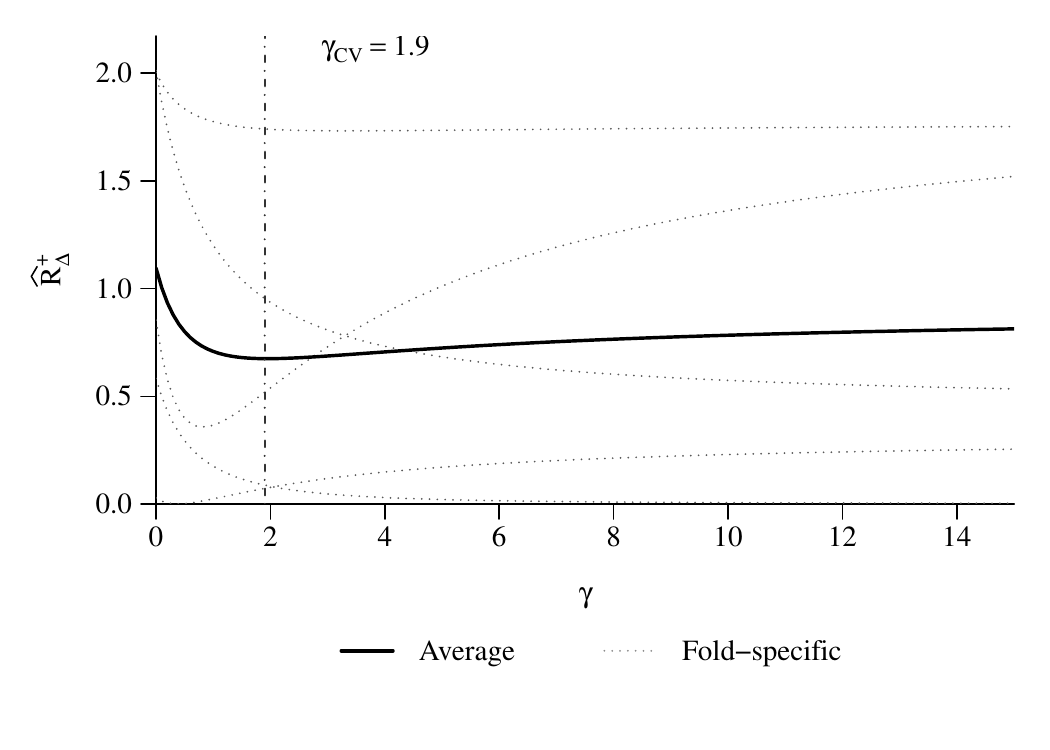}
        \caption*{\textbf{(b)} Cross-validation criterion}
    \end{minipage}

    \caption{\textbf{(a)} The solid curves represent the components of \(\widehat{\beta}_{\gamma}\), while the dotted horizontal lines denote the robust reference parameter \(
\widehat{\beta}_{\mathrm{RL}}
=
(1.27,-0.19)^t
\). The shaded regions denote the corresponding \(95\%\) pointwise Monte Carlo uncertainty bands. The vertical dot-dashed line marks the selected value \(\gamma_{CV}=1.9\), its intersections with the coefficient paths correspond to the components of the selected estimator \(\widehat{\beta}_{CR}\).
    \textbf{(b)} Fold-specific and average values of \(\mathbb{R}_{\Delta}^{+}(\widehat{\beta}_{\gamma})\) as a function of \(\gamma\). The vertical dot-dashed line marks the minimizing value \(\gamma_{CV}=1.9\).}
    \label{fig:exampleC_results}
\end{figure}

Figure~\ref{fig:exampleC_results} (a) shows how the two estimated coefficients change as \(\gamma\) increases. Both paths move quickly towards the robust reference parameter \(\widehat\beta_{RL}\), with most of the variation occurring for small values of \(\gamma\). The pointwise Monte Carlo uncertainty bands widen and are asymmetric around the estimated paths, reflecting the non-symmetric finite-sample
variability observed in this nonlinear example. After this initial adjustment, the coefficients remain nearly constant. The selected value \(\gamma_{CV}=1.9\) identifies a point at which the estimator has already gained substantial stability across the two environments, without moving to the limiting robust solution.

\tikzset{
  causalnode/.style={
    font=\normalsize,
    inner sep=3pt
  },
  causalarrow/.style={
    -{Latex},
    thick,
    black
  }
}

\begin{figure}[t]
\centering

\begin{minipage}[t]{0.48\textwidth}
\vspace{0pt}
\centering

\textbf{Environment $(e_1)$}\\[8pt]

\begin{tikzpicture}
  \node[causalnode] (X11) at (-1.6,0) {$X_1$};
  \node[causalnode] (Y1)  at (0,0)    {$Y$};
  \node[causalnode] (X21) at (1.6,0)  {$X_2$};

  \draw[causalarrow] (X11) -- (Y1);
  \draw[causalarrow] (Y1) -- (X21);
\end{tikzpicture}

\[
\begin{gathered}
X_1^{e_1} \sim \operatorname{Exp}(1),\\
Y^{e_1} \sim \operatorname{Poisson}(X_1^{e_1}),\\
X_2^{e_1} = -Y^{e_1}+\varepsilon_2^{e_1}.
\end{gathered}
\]

\end{minipage}
\hfill
\begin{minipage}[t]{0.48\textwidth}
\vspace{0pt}
\centering

\textbf{Environment $(e_2)$}\\[8pt]

\begin{tikzpicture}
  \node[causalnode] (X12) at (-1.6,0) {$X_1$};
  \node[causalnode] (Y2)  at (0,0)    {$Y$};
  \node[causalnode] (X22) at (1.6,0)  {$X_2$};

  \draw[causalarrow] (X12) -- (Y2);
  \draw[causalarrow] (X22) -- (Y2);
\end{tikzpicture}

\[
\begin{gathered}
X_1^{e_2} \sim \operatorname{Poisson}(1),\\
X_2^{e_2} \sim \operatorname{Exp}(1),\\
Y^{e_2} =
\log\!\left(1+X_1^{e_2}+X_2^{e_2}\right)
+\varepsilon_Y^{e_2}.
\end{gathered}
\]

\end{minipage}

\caption{Causal graphs and corresponding data-generating models for
environments $(e_1)$ and $(e_2)$. In $(e_1)$, $X_1$ affects $Y$,
which subsequently affects $X_2$. In $(e_2)$, both $X_1$ and $X_2$
determine $Y$. The noise variables satisfy
$\varepsilon_2^{e_1}\sim\operatorname{Unif}(-0.5,0.5)$ and
$\varepsilon_Y^{e_2}\sim\mathcal{N}(0,1)$, and are independent of the
corresponding exogenous covariates.}
\label{fig:exampleB_models}
\end{figure}

\section{Causality as prediction limit}
\label{sec:causality_prediction_limit}

We now discuss when the robust population limit of the causal regularization path admits a causal interpretation. The robust equilibrium parameter of the previous section is purely predictive and does not require a causal model. A causal interpretation requires additional assumptions linking this regularizer limit parameter to the response mechanism of \(Y\).

\subsection{Structural causal model and interventions}

We introduce the structural causal model (SCM) notation used for the causal interpretation. This allows us to distinguish interventions on the covariates from interventions on the target.

\begin{definition}[Structural causal model with interventions]
A structural causal model (SCM) is a triple
\[
(\mathcal G,f,\mathcal I),
\]
where \(\mathcal G=(V,E)\) is a directed acyclic graph, \(f=\{f_v\}_{v\in V}\) is a set of structural functions and \(\mathcal I\) is a class of interventions.
For each node \(v\in V\), the structural function is a measurable map
\[
f_v:\mathcal X_{CP(v)}\times [0,1]
\longrightarrow
\mathcal X_v,
\]
where \(CP(v)\) denotes the set of parents of \(v\) in \(\mathcal G\). 
In the absence of interventions, each node is generated structurally as
\[
X_v=f_v(X_{CP(v)},U_v),
\qquad v\in V\setminus V_0.
\]
where the exogenous variable 
\(
U_v\sim \operatorname{Unif}(0,1).
\)
An intervention is specified by a subset \(V_0\subseteq V\) and a distribution \(P^{V_0}\in\mathcal I\) on \(X_{V_0}\). Under the intervention \(P^{V_0}\), the intervened variables are generated as
\[
X_{V_0}\sim P^{V_0}.
\]
If \(V_0=\emptyset\), then all variables are generated structurally.
\end{definition}

In the setting of interest, we write \(V=\{X_1,\ldots,X_p,Y\}\) as the collection of covariates and target. Each environment \(e\) corresponds to an intervention \(P^{V_0}_e\in\mathcal I\) and induces a joint distribution \(P_e^{X,Y}\). We denote by
\[
\mathcal P_{\mathcal I}
=
\left\{
P_e^{X,Y} : P^{V_0}_e\in\mathcal I
\right\}
\]
the class of distributions generated by the admissible interventions with \(\mathcal P_{\mathcal I}\subseteq\mathcal C\). For the causal interpretation developed below, we restrict attention to environments that may perturb the covariate system \(X^e\) and alter the conditional distribution of the response, while preserving its conditional mean given the causal covariates \(X^{e,cp}\).

\begin{example}{\bf Structural Equation Model with additive shifts}

A classical example of a SCM with interventions is a linear structural equation model (SEM) with additive shifts. 
\begin{definition}
A structural equation model with shift interventions is defined as the tuple
$(B,{F}_{\varepsilon},\mathcal{I})$, where $B$ is the
$(p+1)\times(p+1)$ structural coefficient matrix corresponding to an underlying
DAG, ${F}_{\varepsilon}$ is a set of $p+1$ distribution functions for
the structural noise variables, and $\mathcal{I}$ is a set of
$(p+1)$-dimensional shift distributions. For each
$\mathbb{P}_{A^e}\in\mathcal{I}$, this induces in environment $e$ the system
\[
\begin{pmatrix}
Y^e\\
{X}^e
\end{pmatrix}
=
B
\begin{pmatrix}
Y^e\\
{X}^e
\end{pmatrix}
+
{A}^e
+
{\varepsilon}^e,
\]
where ${A}^e\sim\mathbb{P}_{A^e}$ and
$\varepsilon_v^e=F_{\varepsilon_v}^{-1}(U_v^e)$ for i.i.d.\
$U_v^e\sim\operatorname{Unif}(0,1)$, $v\in V$. Furthermore, we assume that
the shifts ${A}^e$ and the structural noise variables
${\varepsilon}^e$ are uncorrelated. 
\(
(I-B)^{-1}
\left({A}^e+{\varepsilon}^e\right).
\)
\end{definition}
We partition $B$ as
\[
B=
\begin{pmatrix}
0 & \beta_{\mathrm{CP}}^{t}\\
\beta_{\mathrm{CH}} & \beta_X
\end{pmatrix}.
\]
Here, $\beta_X$ describes the relations among the covariates,
$\beta_{\mathrm{CH}}$ describes the effects of $Y$ on its descendants
contained in $\boldsymbol{X}$, and $\beta_{\mathrm{CP}}$ contains the
coefficients associated with the causal parents of $Y$.

In the special Gaussian case, the structural assignment for each node \(v\in V\) may equivalently be represented through the inverse probability
transform. Let
\[
U_v^e\sim\operatorname{Unif}(0,1),
\]
and define
\[
X_v^e
=
F^{-1}_{\mu_v^e,\sigma_v^2}(U_v^e),
\]
where \(F_{\mu_v^e,\sigma_v^2}\) denotes the cumulative distribution function of \(\mathcal N(\mu_v^e,\sigma_v^2)\), with
\[
\mu_v^e
=
(X_{CP(v)}^e)^t\beta_v+A_v^e.
\]

\end{example}
The results developed below are not restricted to Gaussianity or additive shifts. This model only provides a familiar setting in which the causal assumptions can be illustrated explicitly. The subsequent theorem is instead formulated in terms of conditional-mean invariance and inner-product invariance.

\begin{theorem}[Causal limit]
\label{theorem:causal}
Let \(\operatorname{CP}(Y)\) denote the set of direct causes of \(Y\) and let \(X^{e}_{CP}\) the covariates relative to the causal parents. Let \(\beta_{CP}\in\mathbb R^p\) be supported only on the causal parent set, that is,
\[
(\beta_{CP})_j=0
\qquad
\text{for all } j\notin CP(Y).
\]
If the first conditional moment of \(Y\) is linear in the causal covariates and invariant across environments:
\begin{equation}
\mathbb E[Y^e\mid X^{e}_{CP}]
=
(X^e)^t\beta_{CP},
\qquad
 \forall e \text{ such that } P_e^{X,Y}\in\mathcal{P}_I.
\label{eq:first_moment_linearity}
\end{equation}
Define the causal residual
\[
r_{CP}^e
=
Y^e-(X^e)^t\beta_{CP}.
\]
Assume moreover that the full inner-product invariance condition holds, namely there exists \(c\in\mathbb R^p\) such that
\begin{equation}
\mathbb E[X^e r_{CP}^e]=c,
\qquad
\forall e\in\mathcal{P}_I.
\label{eq:full_inner_product_invariance}
\end{equation}
If \(G_\Delta\) is non-singular, then the stationary point \(\beta^\star\) of \(R_\Delta\) is unique and, since the population robust limit estimator \(\beta_{RL}\) coincides with the causal parent parameter, we obtain
\[
\beta_{RL}=\beta^\star=\beta_{CP}.
\]

\end{theorem}

A detailed proof can be found in Appendix~\ref{appendix:causal_theorem}.

\begin{corollary}
\label{cor:causal_invariance}

Under the assumptions of Theorem~\ref{theorem:causal}, for every pair of environments satisfying
\(
P_{e_1}^{X,Y},P_{e_2}^{X,Y}\in\mathcal{P}_I,
\)
the causal parameter satisfies both gradient invariance and convexified-risk invariance.
\[
\nabla R_{e_1}(\beta_{CP})
=
\nabla R_{e_2}(\beta_{CP}), \quad R_\Delta^+(\beta_{CP})=0.
\]

\end{corollary}
The mean invariance assumption stated in Equation~\eqref{eq:first_moment_linearity} allows the environments to perturb the distribution of the covariates while requiring the conditional mean mechanism generating \(Y\) from its direct causes to remain unchanged. 
It is weaker than the invariance condition commonly used in invariant causal prediction, which includes the entire conditional distribution \(\mathcal L(Y^e\mid X^{e}_{CP})\) \citep{peters2015causalinferenceusinginvariant} rather than only its conditional mean.

The following proposition shows that mean invariance implies that the residual of the target is conditionally centered and orthogonal to the causal covariates.
\begin{proposition}
Let \(P_e^{X,Y}\in\mathcal{P}_I\). Then the residual \(r_{CP}^e= Y^e-(X^e)^t\beta_{CP}\) satisfies 
\begin{equation*} \mathbb E[r_{CP}^e\mid X^{e}_{CP}]=0 \quad\text{and}\quad \mathbb E[X^{e}_{CP}r_{CP}^e]=0. 
\end{equation*} 
\label{prop:causal}
\end{proposition}

\begin{proof} From Equation~\eqref{eq:full_inner_product_invariance}, 
\begin{align*} \mathbb E[r_{CP}^e\mid X^{e}_{CP}] &= \mathbb E\!\left[ Y^e-(X^e)^t\beta_{{CP}} \mid X^{e}_{CP} \right] \nonumber\\ &= \mathbb E[Y^e\mid X^{e}_{CP}] - (X^e)^t\beta_{{CP}} \nonumber\\ &= 0. \end{align*} 
Taking expectations and applying the law of iterated expectations yields 
\begin{align*} \mathbb E[X^{e}_{CP}r_{CP}^e ] &= \mathbb E\!\left[ X^{e}_{CP} \mathbb E[r_{CP}^e\mid X^{e}_{CP}] \right] \nonumber\\ &= 0. 
\end{align*} 

\end{proof}
However, mean invariance (Equation ~\eqref{eq:first_moment_linearity}) does not imply that the residual is orthogonal to all the components of \(X^e\). Non-causal variables may remain correlated with the residual because of confounding, proxy relations, descendants of \(Y\), or other stable features of the joint distribution. 
The causal parameter is identified as the equilibrium solution induced by the invariance relations across environments, extending the moment-based perspective of causal Dantzig \citep{rothenhausler2019causal}. Consequently, the regularization path interpolates between the pooled OLS solution, obtained for \(\gamma=0\) and the causal parameter \(\beta_{CP}\), approached as \(\gamma \to \infty\).

The preceding result is a population identification statement. As shown in
Section~\ref{subsec:finite-sample}, the finite-sample estimator retains the same form introduced for robust prediction with the stationary endpoint of the regularization path now encoding a causal interpretation.
The cross-validation procedure described in
Section~\ref{sec:cross_validation} in the causal setting selects a finite-sample estimator along a path directed
towards the causal parameter. When \(\gamma=0\), the OLS estimator may
exploit stable predictive associations that are not causal and may consequently be biased with respect to \(\beta_{CP}\), although it is generally estimated with relatively low variability. As \(\gamma\) increases, the estimator places more emphasis on the differences between environments and moves towards the causal solution. This can reduce its bias with respect to \(\beta_{CP}\), but may increase its sampling variability, particularly when the moment differences are weak. 

The next example shows that the same invariance argument extends to an additive-shift model with latent confounding.
\begin{example}{\bf Additive shifts with confounding}
Consider the linear additive-shift SEM where the interventions do not affect \(Y^e\) or \(H^e\) directly. The vector \(X^e\in\mathbb R^p\) represents the observed covariates. There are latent variables represented by \(H^e\in\mathbb R^q\). The variable \(Y^e\) is the response within a specific environment \(e\). 

The structural equations can be written as
\[
\begin{bmatrix}
Y^e\\
X^e\\
H^e
\end{bmatrix}
=
B
\begin{bmatrix}
Y^e\\
X^e\\
H^e
\end{bmatrix}
+
\begin{bmatrix}
\varepsilon_Y^e\\
\varepsilon_X^e\\
\varepsilon_H^e
\end{bmatrix}
+
\begin{bmatrix}
0\\
A_X^e\\
0
\end{bmatrix},
\]
where \(B\) contains the structural relation coefficients among \(Y^e\), \(X^e\) and \(H^e\), specifically: 
\[B=
\begin{pmatrix}
B_{(Y,X),(Y,X)} & B_{(Y,X),H}\\
B_{H,(Y,X)}     & B_{HH}
\end{pmatrix}.
\]
\(A_X^e\) is the environment-specific additive intervention on \(X^e\).

We assume that the distribution of the structural noise vector
\(
\varepsilon^e
=
\begin{pmatrix}
\varepsilon_Y^e\\
\varepsilon_X^e\\
\varepsilon_H^e
\end{pmatrix}
\)
is invariant across environments and satisfies
\(\mathbb E[\varepsilon^e]=0\). In addition, we assume that the intervention
\(A_X^e\) is uncorrelated with the structural noise vector:
\[
\operatorname{Cov}(A_X^e,\varepsilon^e)=0
\qquad\text{for every environment }e.
\]
By marginalizing the confounder \(H^e\), the structural equation model can be written as: 

\[
\begin{pmatrix}
Y^e\\
X^e
\end{pmatrix}
=
\widetilde B
\begin{pmatrix}
Y^e\\
X^e
\end{pmatrix}
+
\begin{pmatrix}
\eta_Y^e\\
\eta_X^e
\end{pmatrix}
+
\begin{pmatrix}
0\\
A_X^e
\end{pmatrix}.
\]
Equivalently, 
\begin{align}
Y^e
&=
(X^e)^t\widetilde\beta_{CP}+\eta_Y^e,
\label{eq:marginalized-Y}\\
X^e
&=
\widetilde\beta_{CH}Y^e
+\widetilde \beta_{XX}X^e
+\eta_X^e+A_X^e,
\label{eq:marginalized-X}
\end{align}
where
\[
\widetilde B
=
B_{(Y,X),(Y,X)}
+
B_{(Y,X),H}
(I-B_{HH})^{-1}
B_{H,(Y,X)}
=
\begin{pmatrix}
0 & \widetilde\beta_{CP}^{t}\\
\widetilde\beta_{CH} & \widetilde \beta_{X}
\end{pmatrix}
\]
and
\[
\begin{pmatrix}
\eta_Y^e\\
\eta_X^e
\end{pmatrix}
=
\begin{pmatrix}
\varepsilon_Y^e\\
\varepsilon_X^e
\end{pmatrix}
+
B_{(Y,X),H}
(I-B_{HH})^{-1}\varepsilon_H^e.
\]

The residual evaluated at \(\widetilde\beta_{CP}\) is
\[
r_{CP}^e
:=
Y^e-(X^e)^t\widetilde\beta_{CP}
=
\eta_Y^e.
\]

The distribution of \(\eta^e\) is obtained through a linear transformation of \(\varepsilon^e\). For this reason, its distribution is the same in every environments, \(\mathbb E[\eta^e]\) is zero and
\[
\operatorname{Cov}(A_X^e,\eta_Y^e)=0.
\]
As a result of this relationship,
\[
\mathbb E[A_X^e\eta_Y^e]=0.
\]

If
\[
\mathbb E[\eta_Y^e\mid X_{\widetilde CP}^e]=0,
\]
then
\[
\mathbb E[Y^e\mid X_{ \widetilde CP}^e]
=
(X^e)^t\widetilde \beta_{CP},
\qquad e \mid P^e \in\mathcal C,
\]
we have shown that the conditional-mean of the target given its parents in the marginalized model is invariant across environments.

To verify inner-product invariance, substitute
Equation~\eqref{eq:marginalized-Y} into
Equation~\eqref{eq:marginalized-X}. We obtain
\[
\left(
I-\widetilde \beta_{X}
-\widetilde\beta_{CH}\widetilde\beta_{CP}^{t}
\right)X^e
=
\eta_X^e+\widetilde\beta_{CH}\eta_Y^e+A_X^e.
\]
Define \(C\) as:
\[
C
:=
\left(
I-\widetilde \beta_{X}
-\widetilde\beta_{CH}\widetilde\beta_{CP}^{t}
\right)^{-1}
\]
Since the structural coefficients are invariant across
environments, \(C\) does not depend on \(e\). It follows that
\[
X^e
=
C\left(
\eta_X^e+\widetilde\beta_{CH}\eta_Y^e+A_X^e
\right).
\]

Using \(r_{CP}^e=\eta_Y^e\), we then have
\[
\begin{aligned}
\mathbb E\!\left[X^e r_{CP}^e\right]
&=
\mathbb E[X^e\eta_Y^e]\\
&=
C\left\{
\mathbb E[\eta_X^e\eta_Y^e]
+
\widetilde\beta_{CH}\mathbb E[(\eta_Y^e)^2]
+
\mathbb E[A_X^e\eta_Y^e]
\right\}\\
&=
C\left\{
\mathbb E[\eta_X\eta_Y]
+
\widetilde\beta_{CH}\mathbb E[\eta_Y^2]
\right\}\\
&=:c.
\end{aligned}
\]
The vector \(c\) does not depend on the environment because \(C\) and
\(\widetilde\beta_{CH}\) are environment-invariant, the distribution of
\(\eta^e\) is invariant and
\(\mathbb E[A_X^e\eta_Y^e]=0\) for every \(e\). The marginalized SEM satisfies inner-product invariance at
\(\widetilde\beta_{CP}\):
\[
\mathbb E\!\left[
X^{e_1}r_{CP}^{e_1}
\right]
=
\mathbb E\!\left[
X^{e_2}r_{CP}^{e_2}
\right]
=
c
\]
for every pair of environments \(e_1\) and \(e_2\).

Moreover, since
\(r_{CP}^e=\eta_Y^e\) and the distribution of
\(\eta_Y^e\) is invariant across environments,
\[
\begin{aligned}
R_e(\widetilde\beta_{CP})
&=
\mathbb E\!\left[
\left(
Y^e-(X^e)^t\widetilde\beta_{CP}
\right)^2
\right]\\
&=
\mathbb E[(\eta_Y^e)^2].
\end{aligned}
\]
The prediction risk at \(\widetilde\beta_{CP}\) is also
invariant:
\[
R_{e_1}(\widetilde\beta_{CP})
=
R_{e_2}(\widetilde\beta_{CP})
\]
for every pair \(e_1,e_2\) in the considered class of environments.
The property of the risk invariant relates only to the special linear SEM with additive shifts. Under the previously stated assumptions, inner product and mean invariance at \(\widetilde\beta_{CP}\) are preserved even when we marginalize over the latent confounders.

\end{example}

\subsection{Illustrative examples}

The following simulations consider a specific instance of the general framework introduced above. In particular, we restrict our attention to linear structural equation models with additive shifts. This setting imposes more structure than the general class and, as shown in the previous section, it implies not only mean and inner-product invariance, but also invariance of the risk evaluated at the causal parameter.
We consider two shifted environments, \(e_1\) and \(e_2\). In the first example in section~\ref{subsec:exc}, we study the aligned case, in which the shift in the second environment is at least as variable as the shift in the first while in section~\ref{subsec:exD}, we relax this ordering and consider the general non-aligned setting.

\subsubsection{The aligned case}
\label{subsec:exc}
The setting considered is a direct generalization of causal regularization \citep{kania2025causalregularizationtradeoffinsample}: we work with two shifted environments, \({e_1}\) and \({e_2}\), and assume that the first is at least as variable as the second, i.e.,  
\[
\mathbb{E}[{A_1}{A_1}^t] \succeq \mathbb{E}[{A_2}{A_2}^t]
\] 
Under this assumption, the second moments of the corresponding covariates are likewise ordered in variability, that is,

\[
\mathbb{E}[{X_{A_1}}{X_{A_1}}^t] \succeq \mathbb{E}[{X_{A_2}}{X_{A_2}}^t]
\]
This ordering guarantees that the risk difference remains convex, irrespective of the value of \(\beta\), exactly as in the original causal regularization setting. Moreover, ensuring that the corresponding difference of the second-moment matrices is positive semidefinite is sufficient to guarantee convexity of \( R_\Delta \) when it is twice differentiable.

Consider a scenario in which we have two environments. Specifically, \(A_1\) influences both \(X_1\) and \(X_2\) while \(A_2\) only \(X_2\), Figure \ref{fig:two-environments_2}. 
\begin{figure}[t]
\centering

\tikzset{
  causalnode/.style={
    font=\large,
    inner sep=2pt
  },
  observational/.style={
    -{Stealth[scale=1.1]},
    semithick,
    black
  },
  intervention/.style={
    -{Stealth[scale=1.1]},
    semithick,
    dashed,
    red!80!black
  },
  anchornode/.style={
    font=\large,
    inner sep=2pt,
    text=red!80!black
  }
}

\begin{minipage}[t]{0.48\textwidth}
\vspace{0pt}
\centering

\textbf{Environment $(e_1)$}\\[6pt]

\begin{tikzpicture}
  \node[causalnode] (X1a) at (-1.6,0) {$X_1$};
  \node[causalnode] (Ya)  at (0,0)    {$Y$};
  \node[causalnode] (X2a) at (1.6,0)  {$X_2$};

  \node[anchornode] (A1) at (0,1.1) {$A_1$};

  \draw[observational] (X1a) -- (Ya);
  \draw[observational] (Ya) -- (X2a);

  \draw[intervention] (A1) -- (X1a);
  \draw[intervention] (A1) -- (X2a);
\end{tikzpicture}

\[
\begin{gathered}
X_1^{e_1}
  = \varepsilon_1^{e_1}+a_{11},\\
Y^{e_1}
  = X_1^{e_1}+\varepsilon_Y^{e_1},\\
X_2^{e_1}
  = -Y^{e_1}+a_{12}+\varepsilon_2^{e_1}.
\end{gathered}
\]

\end{minipage}
\hfill
\begin{minipage}[t]{0.48\textwidth}
\vspace{0pt}
\centering

\textbf{Environment $(e_2)$}\\[6pt]

\begin{tikzpicture}
  \node[causalnode] (X1b) at (-1.6,0) {$X_1$};
  \node[causalnode] (Yb)  at (0,0)    {$Y$};
  \node[causalnode] (X2b) at (1.6,0)  {$X_2$};

  \node[anchornode] (A2) at (0,1.1) {$A_2$};

  \draw[observational] (X1b) -- (Yb);
  \draw[observational] (Yb) -- (X2b);

  \draw[intervention] (A2) -- (X2b);
\end{tikzpicture}

\[
\begin{gathered}
X_1^{e_2}
  = \varepsilon_1^{e_2},\\
Y^{e_2}
  = X_1^{e_2}+\varepsilon_Y^{e_2},\\
X_2^{e_2}
  = -Y^{e_2}+a_{22}+\varepsilon_2^{e_2}.
\end{gathered}
\]

\end{minipage}

\caption{Causal graphs and corresponding data-generating models for
environments $(e_1)$ and $(e_2)$. Black arrows represent the invariant
mechanism $X_1\to Y\to X_2$. Red dashed arrows represent the
environment-specific shifts: $A_1$ acts on $X_1$ and $X_2$ in $(e_1)$,
whereas $A_2$ acts on $X_2$ in $(e_2)$. All noise variables are mutually
independent and distributed as $\mathcal{N}(0,1)$.}
\label{fig:two-environments_2}

\end{figure}
We set
\[
A_1 \;=\;
\begin{bmatrix}
a_{11}  \\
a_{12}      \\
0     
\end{bmatrix}, \quad a_{11}  \sim N(0,1) \quad \text{and} \quad a_{12}  \sim N(0,3)
\]
while
\[
A_2 \;=\;
\begin{bmatrix}
0\\
a_{22}\\   
0
\end{bmatrix},  \quad a_{22} \sim N(0,1).
\]
Notice that we construct $\mathbb{E}[{A_1 A_1^t}]$ and $\mathbb{E}[{A_2 A_2^t}]$ such that their difference is positive semi definite, since
\[
\mathbb{E}[{A_1 A_1^t}]=
\begin{bmatrix}
1 & 0 & 0\\
0 & 3 & 0\\   
0 &0 &0 
\end{bmatrix} \succ  \mathbb{E}[{{A_2}{A_2}^t}]=\begin{bmatrix}
0 & 0&0 \\
0 & 1 & 0\\   
0 & 0 & 0
\end{bmatrix} 
\]
Consequently, the difference between the covariance matrices of the covariates is positive semidefinite, \(G_\Delta \succeq 0\) and \(G_\Delta^+=G_\Delta\). The original risk difference is already convex and no convexification is required.

\begin{figure}[t]
    \centering

    \begin{minipage}[t]{0.48\textwidth}
        \centering
         \includegraphics[width=\linewidth]{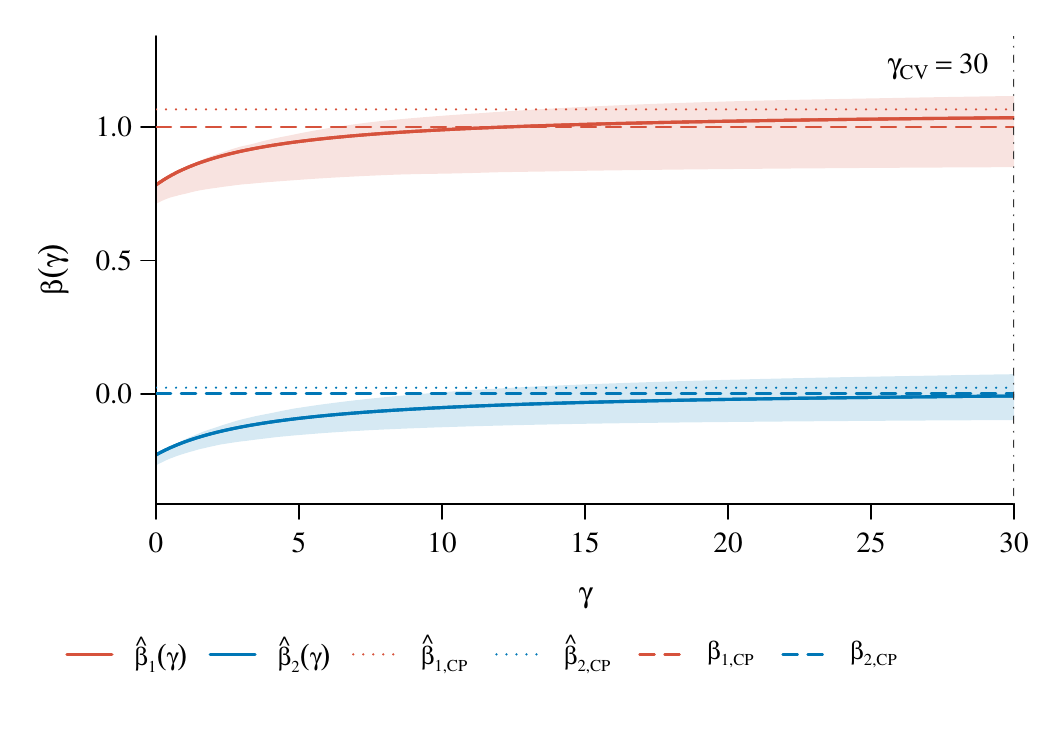}
        \caption*{\textbf{(a)} Regularization path.}
    \end{minipage}
    \hfill
    \begin{minipage}[t]{0.48\textwidth}
        \centering
        \includegraphics[width=\linewidth]{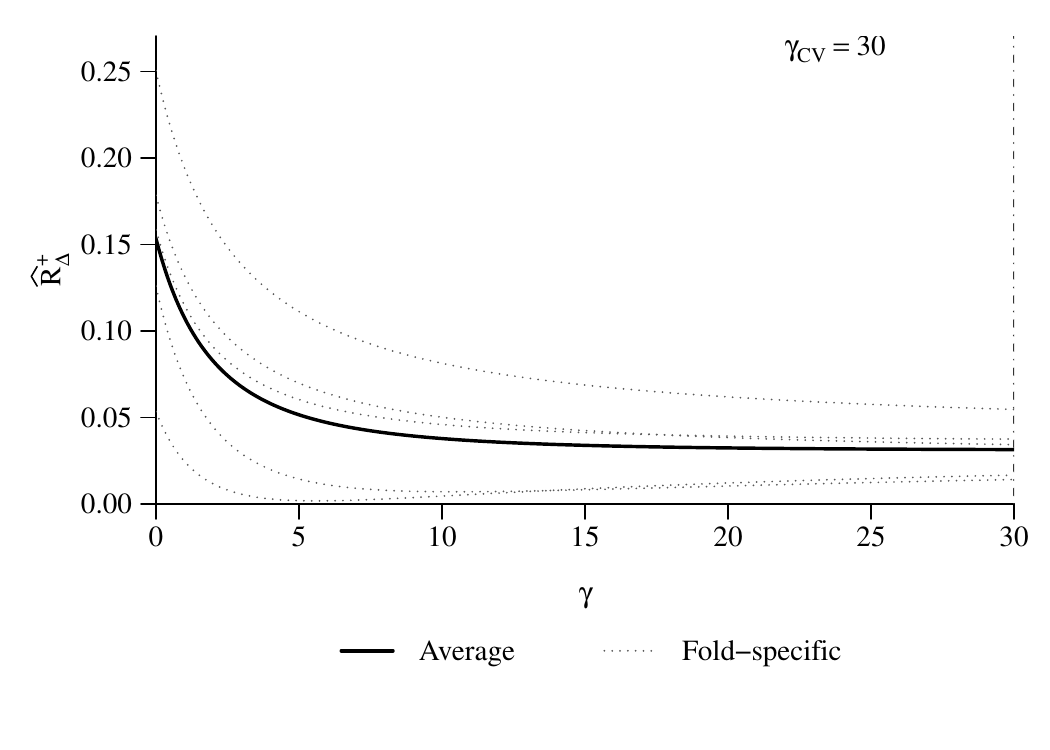}
        \caption*{\textbf{(b)} Cross-validation criterion}
    \end{minipage}
\caption{Results for aligned example. \textbf{(a)} The solid curves represent the components of \(\widehat{\beta}_{\gamma}\). The dotted horizontal lines denote the robust reference parameter \(\widehat{\beta}^{\star}\), while the dashed horizontal lines denote the true causal coefficient \(\beta_{CP}=(1,0)^t\). The vertical line marks the selected value \(\gamma_{CV}=30\). \textbf{(b)} Fold-specific and average values of
\(\mathbb{R}_{\Delta}^{+}(\widehat{\beta}_{\gamma})\). The vertical dot-dashed line marks the minimizing value \(\gamma_{CV}=30\).}
   \label{fig:aligned_results}
\end{figure}

Figure~\ref{fig:aligned_results} summarizes the results for the aligned-shift example. In panel~(a) the graph illustrates how the two estimated coefficients change along the regularization path, together with the predicted causal limit parameter and the true
causal coefficient. Panel~(b) reports the fold-specific and average cross-validation criteria used to select \(\gamma\). 

Cross-validation process identifies the largest number in the set of possible values, \(\gamma_{CV}=30\). Along the regularization path, the data in Figure \ref{fig:aligned_results}(a) indicate that the estimator approaches the true causal coefficient \((1,0)^t\) as gamma increases. The cross-validation criterion drops st a high rate for small values of \(\gamma\) and remains flat, indicating that most of the improvement is obtained in the initial part of the regularization path. The convexified risk differences curve, Figure~\ref{fig:aligned_results} (b), shows a similar decreasing pattern.

\subsubsection{The non-aligned case} 
\label{subsec:exD}

The following simulation is about a more general case where we no longer assume shifts to be aligned. As shown in Figure \ref{fig:two-interventional-environments}, \(X_1\) and \(X_2\) are each influenced separately by \(A_1\) and \(A_2\), respectively, while we currently ignore any direct effect of a shift on \(Y\).
\begin{figure}[t]
\centering

\tikzset{
  causalnode/.style={
    font=\large,
    inner sep=2pt
  },
  anchornode/.style={
    font=\large,
    inner sep=2pt,
    text=red!80!black
  },
  observational/.style={
    -{Stealth[scale=1.1]},
    semithick,
    black
  },
  intervention/.style={
    -{Stealth[scale=1.1]},
    semithick,
    dashed,
    red!80!black
  }
}

\begin{minipage}[t]{0.48\textwidth}
\vspace{0pt}
\centering

\textbf{Environment $(e_1)$}\\[6pt]

\begin{tikzpicture}
  \node[causalnode] (X1a) at (-1.6,0) {$X_1$};
  \node[causalnode] (Ya)  at (0,0)    {$Y$};
  \node[causalnode] (X2a) at (1.6,0)  {$X_2$};

  \node[anchornode] (A1) at (0,1.1) {$A_1$};

  \draw[observational] (X1a) -- (Ya);
  \draw[observational] (Ya) -- (X2a);

  \draw[intervention] (A1) -- (X1a);
\end{tikzpicture}

\[
\begin{gathered}
X_1^{e_1}
  = \varepsilon_1^{e_1}+a_{11},\\
Y^{e_1}
  = X_1^{e_1}+\varepsilon_Y^{e_1},\\
X_2^{e_1}
  = -Y^{e_1}+\varepsilon_2^{e_1}.
\end{gathered}
\]

\end{minipage}
\hfill
\begin{minipage}[t]{0.48\textwidth}
\vspace{0pt}
\centering

\textbf{Environment $(e_2)$}\\[6pt]

\begin{tikzpicture}
  \node[causalnode] (X1b) at (-1.6,0) {$X_1$};
  \node[causalnode] (Yb)  at (0,0)    {$Y$};
  \node[causalnode] (X2b) at (1.6,0)  {$X_2$};

  \node[anchornode] (A2) at (0,1.1) {$A_2$};

  \draw[observational] (X1b) -- (Yb);
  \draw[observational] (Yb) -- (X2b);

  \draw[intervention] (A2) -- (X2b);
\end{tikzpicture}

\[
\begin{gathered}
X_1^{e_2}
  = \varepsilon_1^{e_2},\\
Y^{e_2}
  = X_1^{e_2}+\varepsilon_Y^{e_2},\\
X_2^{e_2}
  = -Y^{e_2}+\varepsilon_2^{e_2}+a_{22}.
\end{gathered}
\]

\end{minipage}

\caption{Causal graphs and corresponding data-generating models for
environments $(e_1)$ and $(e_2)$. Black arrows represent the invariant
mechanism $X_1\to Y\to X_2$. Red dashed arrows represent the
environment-specific shifts: $A_1$ acts on $X_1$ in $(e_1)$, whereas
$A_2$ acts on $X_2$ in $(e_2)$. All noise variables are mutually
independent and distributed as $\mathcal{N}(0,1)$.}
\label{fig:two-interventional-environments}

\end{figure}

The shifts target different covariates and do not interact. Using the
coordinate ordering \((X_1,X_2,Y)\), we set
\[
A_1=
\begin{bmatrix}
a_{11}\\
0\\
0
\end{bmatrix},
\qquad
A_2=
\begin{bmatrix}
0\\
a_{22}\\
0
\end{bmatrix},
\qquad
a_{11}{\sim}\mathcal{N}(3,1),\quad a_{22}{\sim}\mathcal{N}(1,1).
\]

\begin{figure}[t]
    \centering

    \begin{minipage}[t]{0.48\textwidth}
        \centering
        \includegraphics[width=\linewidth]{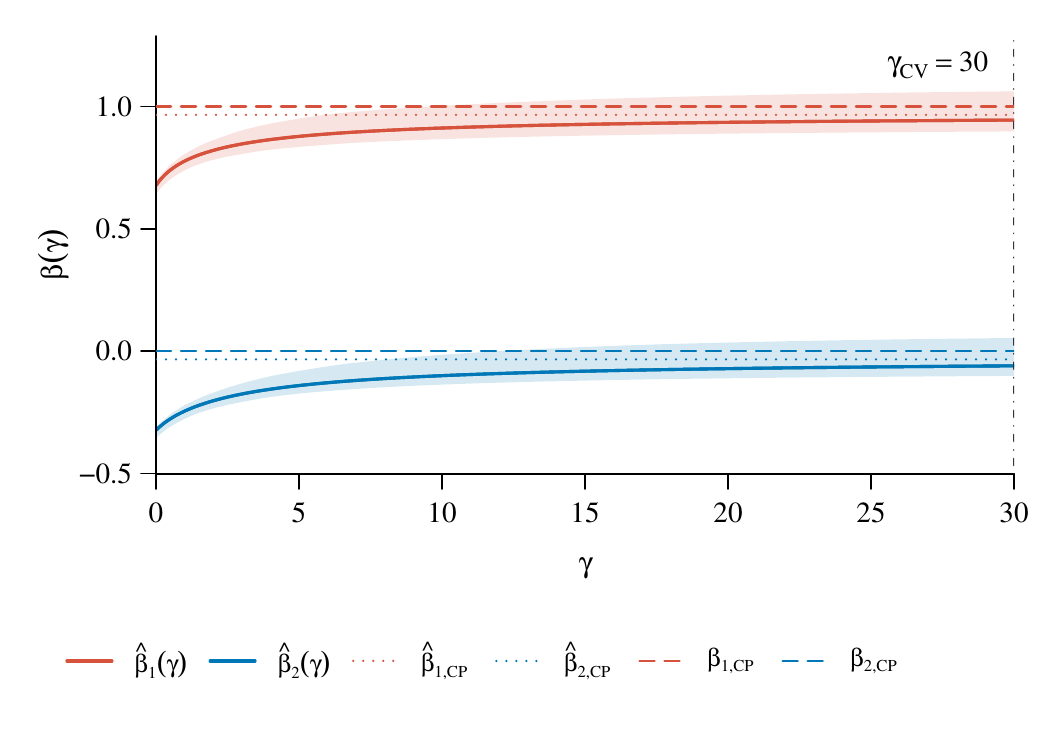}
        \caption*{\textbf{(a)} Regularization path.}
    \end{minipage}
    \hfill
    \begin{minipage}[t]{0.48\textwidth}
        \centering
        \includegraphics[width=\linewidth]{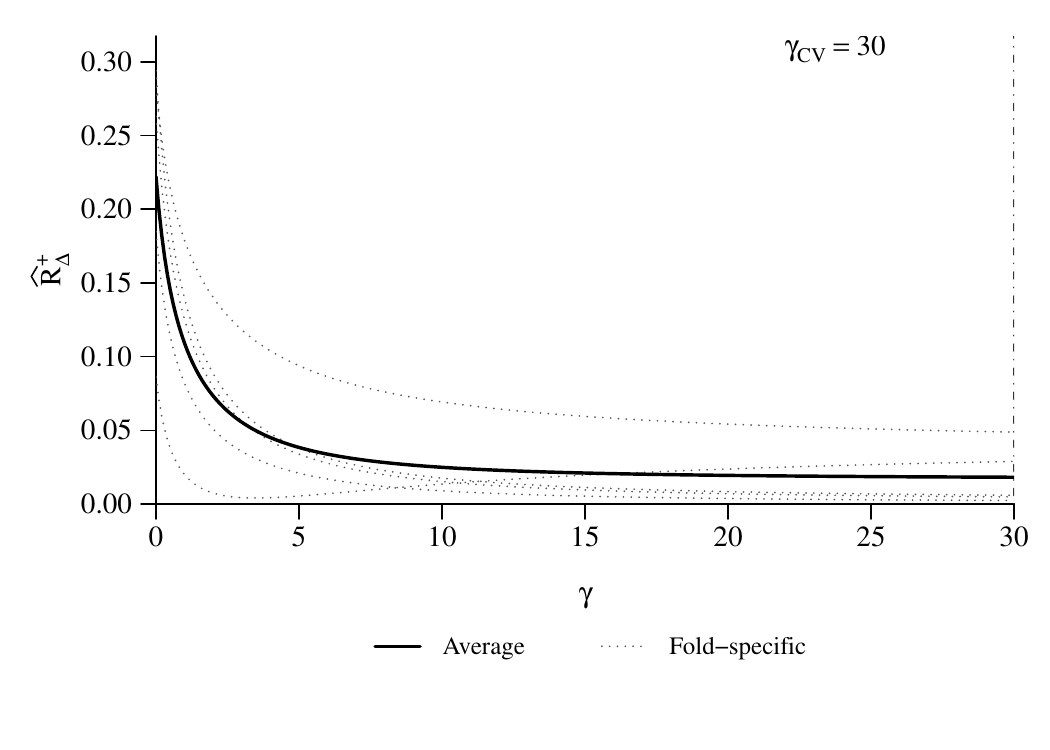}
        \caption*{\textbf{(b)} Cross-validation criterion}
    \end{minipage}
\caption{Results for non-aligned example. \textbf{(a)} The solid curves represent the components of \(\widehat{\beta}_{\gamma}\). The dotted horizontal lines denote the robust reference parameter \(\widehat{\beta}^{\star}\),
while the dashed horizontal lines denote the true causal coefficient \(\beta_{CP}=(1,0)^t\). The vertical line marks the selected value \(\gamma_{CV}=30\). \textbf{(b)} Fold-specific and average values of
\(\mathbb{R}_{\Delta}^{+}(\widehat{\beta}_{\gamma})\) as a function of \(\gamma\). The vertical dot-dashed line marks the minimizing value \(\gamma_{CV}=30\).}
   \label{fig:exampleD_results}
\end{figure}
Figure~\ref{fig:exampleD_results} shows that, as \(\gamma\) increases, the estimated coefficients move toward the true causal parameter \(\beta_{CP}=(1,0)^t\). Cross-validation selects \(\gamma_{CV}=30\), where the average validation risk reaches its minimum. As Simulation C (Section~\ref{subsec:exc}), the selected value coincides with the extreme value of the considered grid. As before, the latter example favours the limiting causal solution. 

\subsection{Comparison with anchor regression}

To compare causal regularization and anchor regression effectively, we analyze both methods within the context of linear structural equation models. So far, we have considered two distinct environments with sample sizes \(n_1\) and \(n_2\). Anchor regression does not require the data from these environments to be fitted as separate regression problems.
We pool all observations into a single design matrix and a single response vector. The total sample size is
\(
n=n_1+n_2.
\)
The two environments remain the systems that generate the data, but their observations are no longer treated as separate datasets. Information about the environment is instead encoded through the anchor variable \(D\).

In principle, anchors are used to represent exogenous shifts affecting the system. Since the actual shifts are not observed in our setting, we use a binary dummy variable as the anchor. This variable indicates the environment to which each observation belongs.

\[
D_i=
\begin{cases}
0,& i\in e_1,\\
1,& i\in e_2.
\end{cases}
\]
The sample mean is \(\bar D =\frac{n_2}{n}.\)
We use the centered environment dummy 
\[
\widetilde{\mathbf{D}}=\mathbf D-\bar D\mathbf{1}_n
\]

The corresponding orthogonal projection onto the space spanned by the centered anchor is:
\[
\mathbf P_{\widetilde{\mathbf{D}}}
=
\widetilde{\mathbf{D}}
\left(
\widetilde{\mathbf{D}}^t\widetilde{\mathbf{D}}
\right)^{-1}
\widetilde{\mathbf{D}}^t.
\]
For a candidate coefficient vector \(\beta\), the pooled residual
vector is defined as 
\(
\mathbf{r}(\beta)
= \mathbf y -\mathbf X \beta.
\)

In the two-environment setting, the squared Euclidean norm of the projection of the residual vector onto the space spanned by the centered binary anchor is
\[
\begin{aligned}
\left\|\mathbf P_{\widetilde{\mathbf{D}}}\mathbf{r}(\beta)\right\|^2
&=
\frac{n_1n_2}{n}
\left(
\bar r_{e_2}(\beta)-\bar r_{e_1}(\beta)
\right)^2,
\end{aligned}
\]
where \(\bar r_{e_1}(\beta)\) and \(\bar r_{e_2}(\beta)\) are the average residuals in the two environments. The anchor penalty controls the difference between the environment-specific average residuals. The detailed derivation of this identity is provided in Appendix~\ref{appendix:ar}.

The anchor-regression estimator is defined by
\[
\widehat\beta_{\mathrm{AR}}(\gamma)
=
\arg\min_\beta
\left\{
\frac{1}{n}
\left\|
(I-\mathbf P_{\widetilde{\mathbf{D}}})\mathbf{r}(\beta)
\right\|^2
+
\frac{\gamma}{n}
\left\|
\mathbf P_{\widetilde{\mathbf{D}}}\mathbf{r}(\beta)
\right\|^2
\right\}.
\]

As $\gamma\rightarrow\infty$, anchor regression solves
\[
\widehat{\beta}_{\mathrm{AR},\infty}
=
\arg\min_{\beta}\|\mathbf y-\mathbf X\beta\|_2^2
\quad\text{subject to}\quad
\widetilde{\mathbf{D}}^t(\mathbf y- \mathbf X\beta)=0.
\]

Under the assumptions adopted throughout the framework, the limiting anchor-regression estimator admits the following closed-form expression:
\[
\widehat{\beta}_{\mathrm{AR},\infty}
=
\widehat{\beta}_{\mathrm{OLS}}
+
(\mathbf X^t \mathbf X)^{-1}\mathbf X^t\widetilde{\mathbf{D}}\,
\frac{
\widetilde{\mathbf{D}}^t \mathbf y
-
\widetilde{\mathbf{D}}^t \mathbf X\widehat{\beta}_{\mathrm{OLS}}
}{
\widetilde{\mathbf{D}}^t \mathbf X
(\mathbf X^t \mathbf X)^{-1}
\mathbf X^t\widetilde{\mathbf{D}}
}
\]
The derivation of the closed-form expression for
\(\widehat\beta_{\mathrm{AR},\infty}\) are reported in
Appendix~\ref{appendix:ar}.
Treating the environment indicator as an anchor is possible, but it does not guarantee that anchor regression converges to the causal coefficient vector. 

\begin{example}
    Using the same setting as in the non-aligned example \ref{subsec:exD}, we compare the regularization paths obtained for causal regularization (CR) and anchor regression (AR). Both methods are calculated without an intercept. Figure~\ref{fig:example-D-CR-AR} displays the estimated coefficients as
they change according to the parameter \(\gamma\).

\begin{figure}[t]
    \centering
    \includegraphics[width=0.5\linewidth]{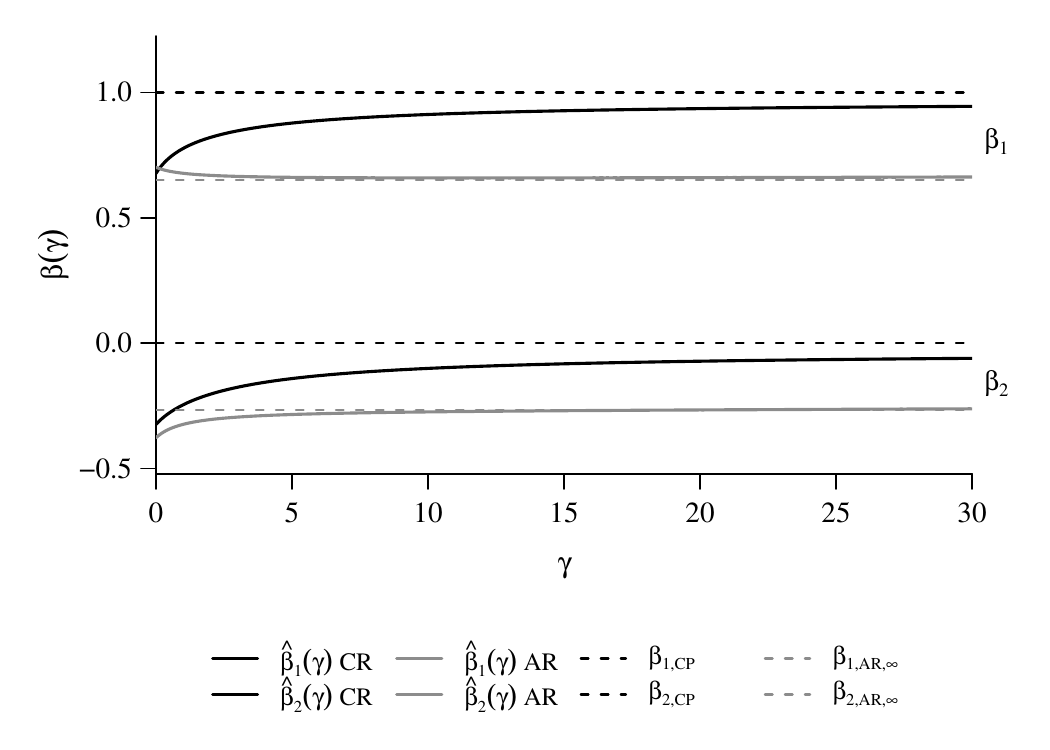}
 \caption{Comparison of the causal regularization (CR) and anchor
regression (AR) paths in the non-aligned setting of
Section~\ref{subsec:exD}. The solid black curves show the CR estimates
\(\widehat{\beta}_{1}(\gamma)_{\mathrm{CR}}\) and
\(\widehat{\beta}_{2}(\gamma)_{\mathrm{CR}}\). The solid grey curves
show the AR estimates
\(\widehat{\beta}_{1}(\gamma)_{\mathrm{AR}}\) and
\(\widehat{\beta}_{2}(\gamma)_{\mathrm{AR}}\), computed using the
centered binary environment anchor. The black dashed lines indicate
the causal coefficients \(\beta_{1,\mathrm{CP}}=1\) and
\(\beta_{2,\mathrm{CP}}=0\). The grey dashed lines indicate the
limiting AR coefficients as \(\gamma\to\infty\).}
    \label{fig:example-D-CR-AR}
\end{figure}
As we can see from the result, causal regularization gets closer to the causal parameter \((\beta_{1,\mathrm{CP}},\beta_{2,\mathrm{CP}}) =(1,0)\). On the other hand, anchor regression does not recover the causal solution in this configuration when \(\gamma\) goes to infinity. The centered binary anchor creates only one restriction, the mean residuals are equal in both environments. This single restriction is insufficient to identify the two-dimensional causal parameter. 

\end{example}

\section{Empirical illustrations}
\label{sec:application}
\subsection{Single-Cell perturbation data from RPE1 cells} 

We evaluate the proposed causal-regularization procedure on the genome-scale Perturb-seq dataset of \cite{replogle2022mapping}. The dataset contains single-cell
RNA-sequencing measurements of cells from the human retinal pigment epithelial cell line hTERT RPE-1 (hereafter, RPE1), collected under different gene interventions. Following the preprocessing approach of ~\cite{chevalley2025large} and the single-cell analysis of \cite{shen2026causality}, we restrict the analysis to ten highly expressed genes, treating one gene as the response variable and the remaining nine as covariates in a linear regression model. The interventional environment pools the cells obtained by separately intervening on four out of the nine predictor genes and contains 1,120 cells. The second environment is the non-targeting and contains
11,485 cells. Consequently, the two training environments differ through interventions acting directly on the observed predictor vector.

\begin{figure}[t]
    \centering
    \begin{subfigure}[t]{0.48\linewidth}
        \centering
        \includegraphics[width=\linewidth]
        {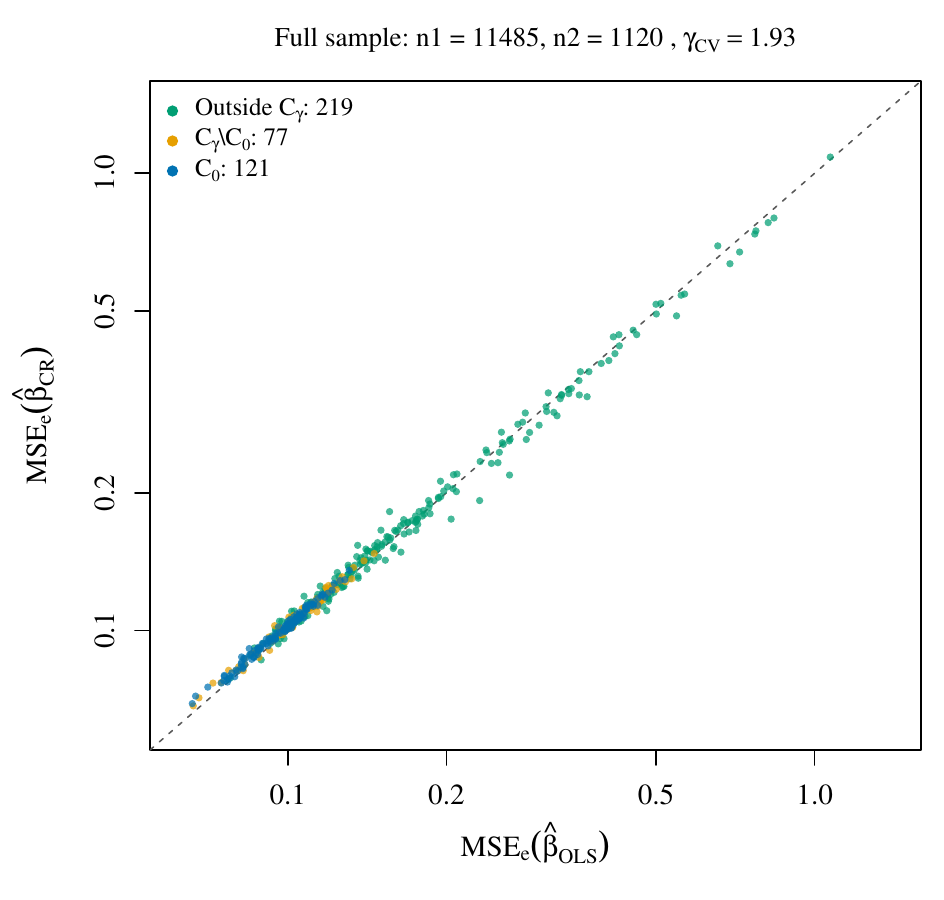}
        \caption{Full training sample: $n_1=11485$ and $n_2=1120$.}
        \label{fig:rpe-scatter-full}
    \end{subfigure}
    \hfill
    \begin{subfigure}[t]{0.48\linewidth}
        \centering
        \includegraphics[width=\linewidth]
        {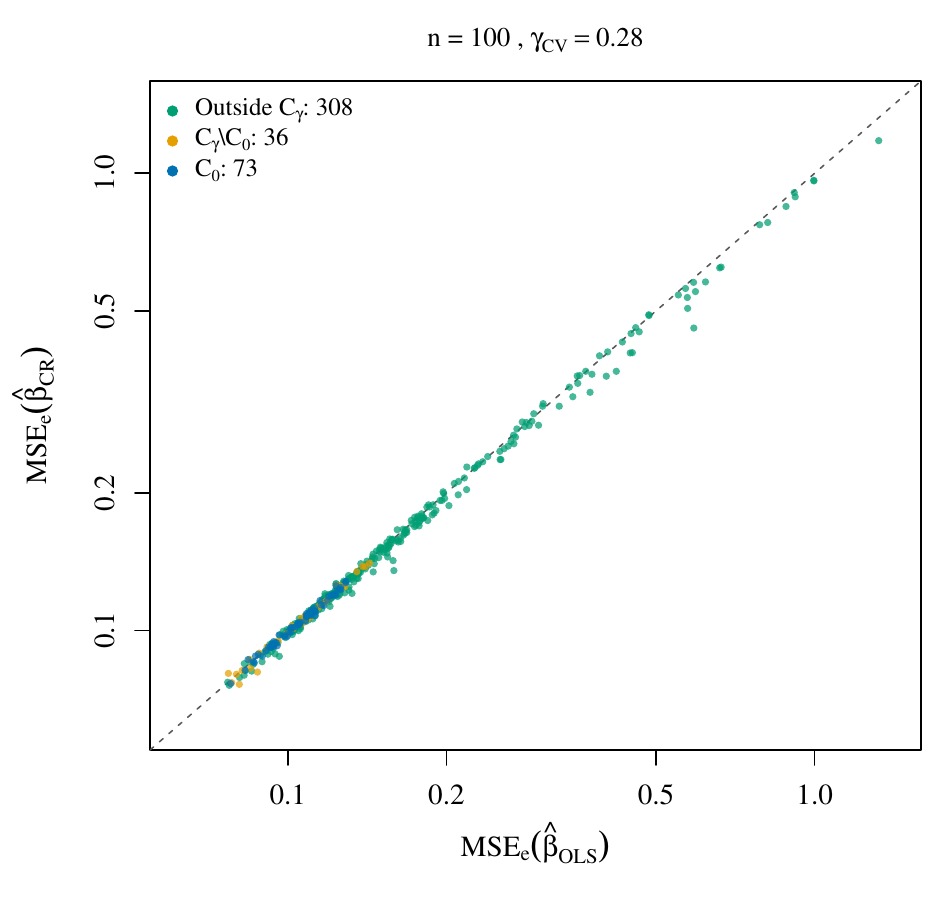}
        \caption{Subsampled training environments: $n_1=n_2=100$.}
        \label{fig:rpe-scatter-n100}
    \end{subfigure}

    \caption{Environment-specific prediction errors of OLS and CR using the complete training samples (left) and subsamples of size $100$ from each training environment (right). Both axes are displayed on a logarithmic scale. Points below the diagonal correspond to test environments in which CR attains a lower MSE than OLS.}
    \label{fig:rpe-scatter-comparison}
\end{figure}

We apply the proposed method on these data. The regularization parameter \(
\gamma_{\mathrm{CV}}=1.93
\) is selected using 10-fold cross-validation applied exclusively to the observed training environments. All model parameters are subsequently re-estimated using the complete training sample. After excluding the environments used for training, 417 previously unseen interventional environments remain available for evaluation. For each test environment, we compute the mean squared prediction error separately for \(\widehat{\beta}_{\mathrm{OLS}}\) and
\(\widehat{\beta}_{\mathrm{CR}}\). Figure~\ref{fig:rpe-scatter-comparison} compares the environment-specific errors using the complete training samples
(a) and subsamples of 100 cells from each training environment
(b). Each point represents one test environment, while the diagonal
indicates equal predictive performance of the two estimators.

Figure~\ref{fig:rpe-scatter-comparison} shows the relationship between the performance of causal regularization and the size of the training sample. With the complete training sample, the cross-validation selects \(\gamma_{\mathrm{CV}}=1.93\). By looking at Figure~\ref{fig:rpe-scatter-comparison}(a), around \(75.3\%\) of the points lie above the diagonal, indicating that OLS produces a lower environment-specific MSE than the estimator chosen by cross validation. When the training data contains many observationa, OLS estimates parameters with high precision, whereas increasing \(\gamma\) gives more weight on the heterogeneity term \(R_\Delta^+\) and may introduce unnecessary variability. Although large values of \(\gamma\) provide greater robustness to distributional shifts, this additional robustness can reduce predictive accuracy in environments that are close to the pooled training distribution.
This interpretation is also consistent with the approximate classification of the test environments based on their second-order similarity to the training environments. In the full-sample analysis, 121 test environments are classified as belonging to \(\mathcal C_0\) and 77 belong to \(\mathcal C_{\gamma_{\mathrm{CV}}}\setminus\mathcal C_0\). There are many conditions in \(\mathcal C_0\), which explains why OLS performs well. 

When the training environments are subsampled to \(n_1=n_2=100\), cross-validation selects the smaller value \(\gamma_{\mathrm{CV}}=0.28\). With the full sample, OLS is already sufficiently accurate. With only 100 observations per environment OLS is
more affected by sampling variability, and moderate regularization can
improve its performance by incorporating information about the
differences between the training environments. This is evident in Figure \ref{fig:rpe-scatter-comparison}(b): only 95 of the 417 test
environments (\(22.8\%\)) lie above the diagonal. The improvement of regularization is particularly visible for test environments with extreme prediction errors, including environments classified as lying outside \(\mathcal C_{\gamma}\). For several of these environments, the cross-validated estimator lies below the diagonal and substantially outperforms OLS. 

\begin{figure}[t]
    \centering
    \begin{subfigure}[t]{0.48\linewidth}
        \centering
        \includegraphics[width=\linewidth]
        {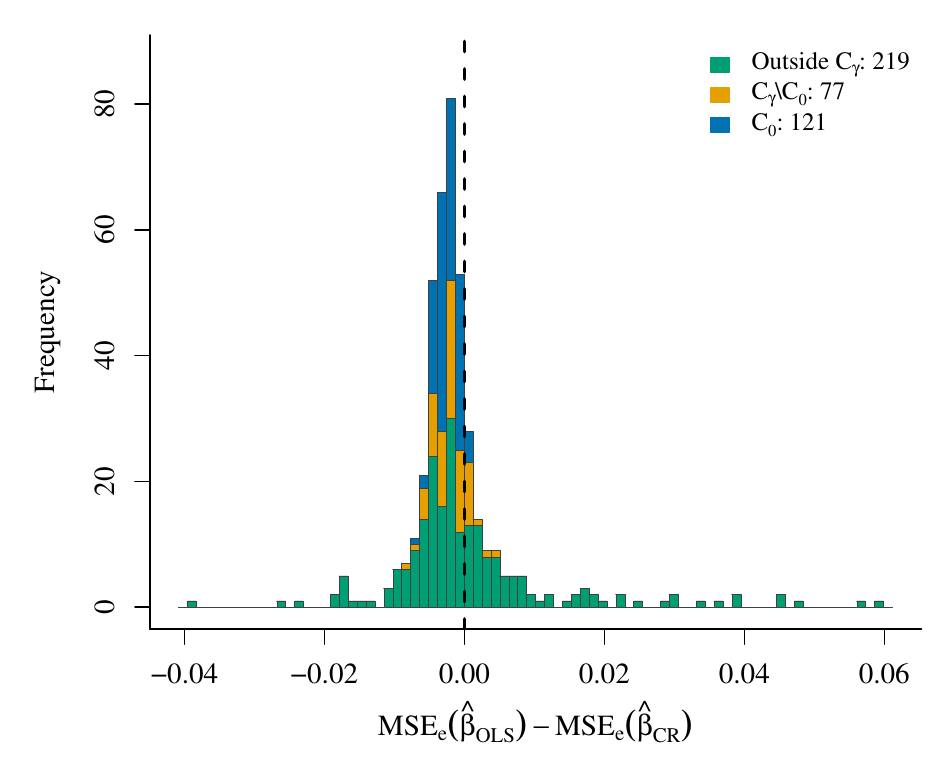}
        \caption{Full training sample: $n_1=11485$ and $n_2=1120$.}
        \label{fig:rpe-hist-full}
    \end{subfigure}
    \hfill
    \begin{subfigure}[t]{0.48\linewidth}
        \centering
        \includegraphics[width=\linewidth]
        {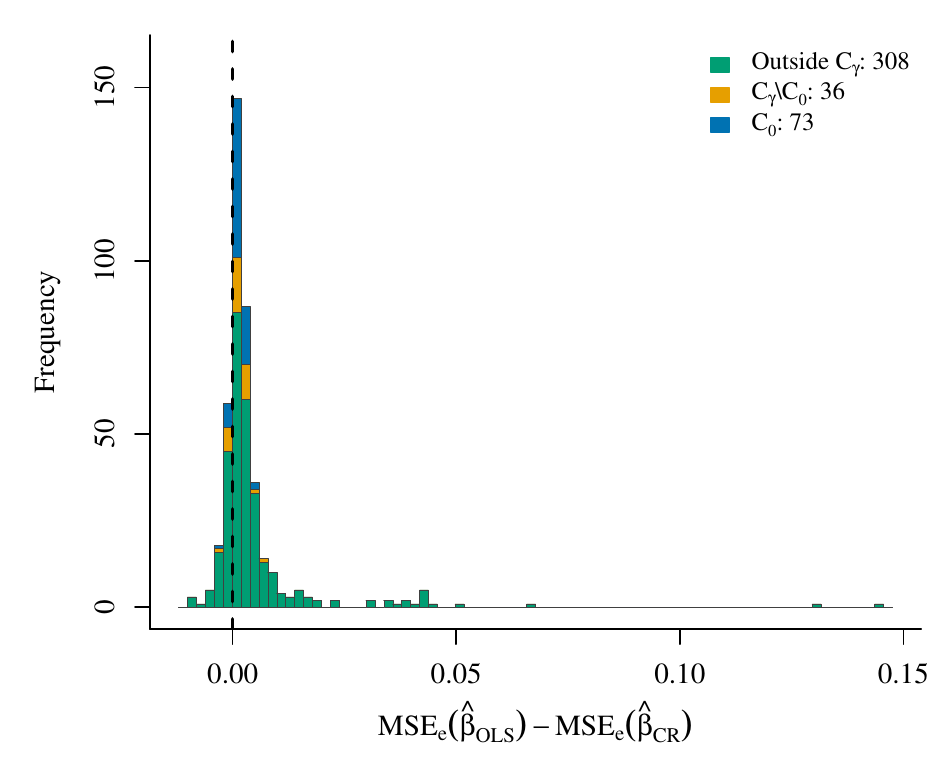}
        \caption{Subsampled training environments: $n_1=n_2=100$.}
        \label{fig:rpe-hist-n100}
    \end{subfigure}

    \caption{Distribution across test environments of $\operatorname{MSE}_e(\widehat\beta_{\mathrm{OLS}})
    -\operatorname{MSE}_e(\widehat\beta_{\mathrm{CV}})$. Positive values indicate that CR attains a lower prediction error than OLS, while negative values favour OLS. The vertical dashed line marks equality between the two methods.}
    \label{fig:rpe-hist-comparison}
\end{figure}
Figure~\ref{fig:rpe-hist-comparison} complements the scatterplots by showing the distribution of the MSE difference across the test environments. With the full training sample, the distribution is concentrated around zero, with its mode slightly on the negative side. OLS performs marginally better than CR in many test environments, although the differences are generally small. The positive right tail shows that CR yields larger improvements in a smaller number of environments.

When the training sample is reduced to \(n_1=n_2=100\), the distribution becomes right-skewed. Most differences remain close to zero, but the positive tail is longer than the negative tail.
The two estimators perform similarly in most test environments, while CR performs substantially better in several environments, reducing some of the largest prediction errors.

\subsection{Causal chamber}

In this section, we investigate the causal behavior of the proposed method using data from the causal chamber~\citep{gamella2025chamber}, a recently developed physical testbed designed for the evaluation and comparison of causal inference methods in controlled but realistic experimental settings. A key advantage of this dataset is that the underlying causal graph is known, a property that is rarely available in real-world applications. We consider data generated by the light tunnel setup. The system contains a configurable light source, several light sensors, local LEDs and two polarizers. These components make it possible to generate different experimental environments by directly intervening in selected variables of the physical system.

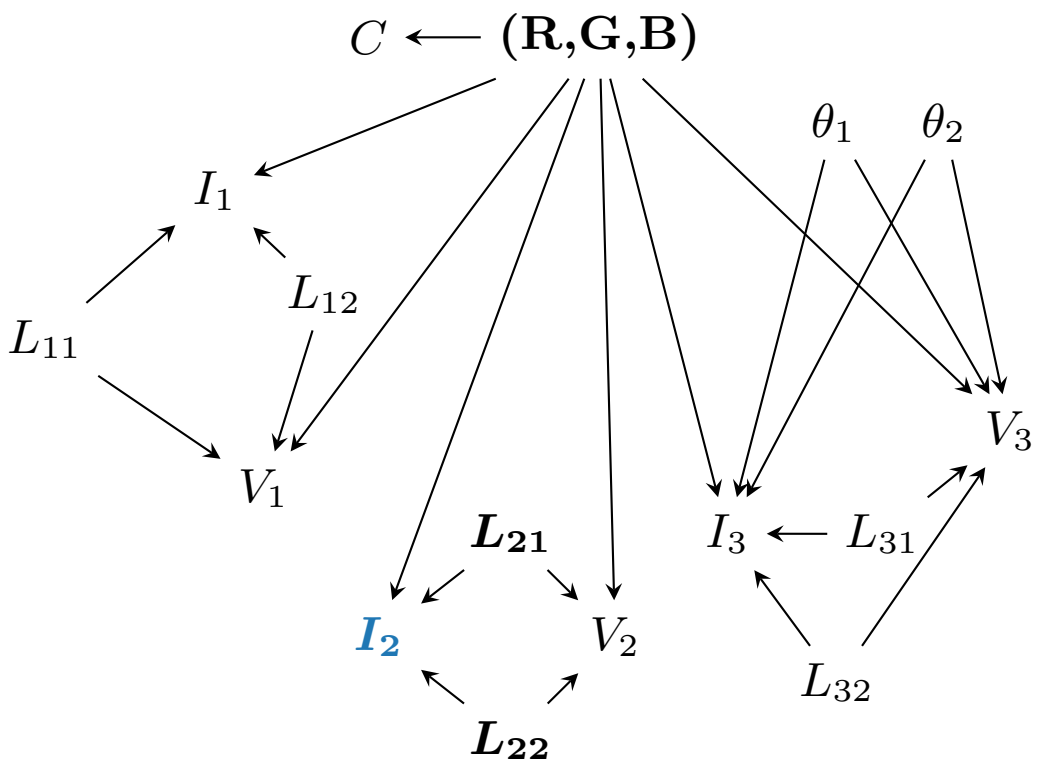
\begin{figure}[t]
\centering
\begin{tikzpicture}[
    scale=1.2,
    node distance=1cm and 1cm,
    every node/.style={draw=none, text centered},
    every path/.style={draw, -stealth}
]

\node (RGB) {\textbf{(R,G,B)}};
\node[left=0.5cm of RGB] (Ctilde) {$C$};

\node[below left=0.5cm and 1.5cm of RGB] (I1tilde) {$I_1$};
\node[below right=0.2cm and 0.1cm of I1tilde] (L12) {$L_{12}$};
\node[below left=0.5cm and 0.5cm of I1tilde] (L11) {$L_{11}$};
\node[below right=0.5cm and 0.8cm of L11] (V1tilde) {$V_1$};

\node[below right=0.5cm and 0.2cm of V1tilde]
    (I2tilde) {\textcolor{blue}{$\boldsymbol{I_2}$}};
\node[above right=0.2cm and 0.2cm of I2tilde]
    (L21) {$\boldsymbol{L_{21}}$};
\node[below right=0.2cm and 0.2cm of I2tilde]
    (L22) {$\boldsymbol{L_{22}}$};
\node[right=1cm of I2tilde] (V2tilde) {$V_2$};


\node[below right=0.05cm and 0.5cm of RGB] (Theta1) {$\theta_1$};
\node[right=0.2cm of Theta1] (Theta2) {$\theta_2$};
\node[above right=0.2cm and 0.2cm of V2tilde] (I3tilde) {$I_3$};
\node[right=0.4cm of I3tilde] (L31) {$L_{31}$};
\node[above right=0.2cm and 0.2cm of L31] (V3tilde) {$V_3$};
\node[below right=0.5cm and 0.1cm of I3tilde] (L32) {$L_{32}$};

\draw (RGB) -- (Ctilde);
\draw (RGB) -- (I1tilde);
\draw (RGB) -- (V1tilde);
\draw (L11) -- (I1tilde);
\draw (L12) -- (I1tilde);
\draw (L11) -- (V1tilde);
\draw (L12) -- (V1tilde);

\draw (RGB) -- (I2tilde);
\draw (RGB) -- (V2tilde);
\draw (L21) -- (I2tilde);
\draw (L21) -- (V2tilde);
\draw (L22) -- (I2tilde);
\draw (L22) -- (V2tilde);

\draw (RGB) -- (I3tilde);
\draw (RGB) -- (V3tilde);
\draw (L31) -- (I3tilde);
\draw (L31) -- (V3tilde);
\draw (L32) -- (I3tilde);
\draw (L32) -- (V3tilde);
\draw (Theta1) -- (I3tilde);
\draw (Theta2) -- (I3tilde);
\draw (Theta1) -- (V3tilde);
\draw (Theta2) -- (V3tilde);

\end{tikzpicture}
\caption{Known causal graph of the causal chamber light tunnel. The response variable $I_2$ is highlighted in blue.}
\label{fig:causalchamb-graph}
\end{figure}

Figure~\ref{fig:causalchamb-graph} shows the known causal graph of the light tunnel. The color settings of the main light source are denoted by $R$, $G$ and $B$, while $C$ denotes the current drawn by the light source. The variables $I_i$ and $V_i$ denote, respectively, the infrared and visible light intensities measured at sensor $i$. The variables $L_{i1}$ and $L_{i2}$ represent the brightness levels of the two LEDs located near sensor $i$ and $\theta_1$ and $\theta_2$ denote the positions of the two polarizers.
For the analysis, we focus on the prediction of the continuous infrared intensity measured at the second sensor, denoted by
\(
Y = I_2
\). According to the known causal graph, the structural direct causes of $I_2$ are $R$, $G$, $B$, $L_{21}$, $L_{22}$.

In the estimation procedure, we retain all 17 numerical covariates. Metadata and diagnostic variables that are not represented in the causal graph are excluded.
The analysis considers \(53\) intervention environments, each containing \(1{,}000\) observations. Every pair is treated as a separate two-environment training problem, resulting in
\(
\binom{53}{2}=1{,}378
\) experiments.
We model each covariate using an additive quadratic spline to account for potentially nonlinear physical relations. The predictors are standardized using their pooled means and standard deviations. 
Let \( X_j\) denote the standardized version of covariate \(j\). For \(x\in\mathbb R\), we use the quadratic spline basis
\[
b(x)
:=
\begin{pmatrix}
x\\
x^2\\
[\max(0,x)]^2
\end{pmatrix}.
\]
The knot is located at zero, allowing the curvature of the fitted relation to differ below and above the pooled average mean. The resulting model contains three basis terms for each of the \(17\) covariates, for a total of \(51\) coefficients.

For each environment pair, we estimate the pooled OLS estimator and the proposed estimator, selecting the regularization parameter by five-fold cross-validation. When \(\widehat\gamma_{\mathrm{CV}}=0\), the proposed estimator coincides with OLS and the two procedures produce identical estimates.
Because each covariate is represented by three spline coefficients, we evaluate its overall contribution by jointly testing these coefficients. For this purpose, we estimate the conditional covariance matrix of each estimator and construct a Hotelling \(T^2\) test for every original covariate. 

We estimate the conditional variance matrices using Proposition~\ref{prop:conditional-variance}, treating the cross-validated value of $\gamma$ as fixed. For OLS, we use the same expression with $\gamma=0$.
We then test the joint null hypothesis
\[
H_{0,j}:
\beta_{\gamma,j1}
=
\beta_{\gamma,j2}
=
\beta_{\gamma,j3}
=
0
\]
using the Hotelling \(T^2\) statistic
\[
T_{\gamma,j}^2
=
\widehat\beta_{\gamma,j}^{t}
\widehat V_{\gamma,j}^{-1}
\widehat\beta_{\gamma,j}.
\]
Under \(H_{0,j}\), the statistic follows a Hotelling \(T^2\)
distribution, which can be expressed in terms of an \(F\)-distribution.
In our implementation, we use
\[
F_{\gamma,j}
=
\frac{T_{\gamma,j}^2}{3},
\]
and compute the \(p\)-value from
\[
F_{\gamma,j}\sim F_{3,\nu}.
\]
where
\[
\nu
=
n_{e_1}+n_{e_2}
-
\operatorname{rank}
\begin{pmatrix}
X^{e_1}\\
X^{e_2}
\end{pmatrix},
\]
with \(n_1=n_2=1000\) and \(\operatorname{rank}\begin{pmatrix}
X^{e_1}\\
X^{e_2}\end{pmatrix}=51\).
We reject \(H_{0,j}\) when the resulting \(p\)-value is smaller than
\(\alpha=0.05\).
The same test is applied to OLS using its coefficient vector
\(\widehat\beta_{\mathrm{OLS}}\) and conditional covariance matrix
\(\operatorname{Var}
\left(
\widehat\beta_{\mathrm{OLS}}\mid X
\right)\). Rejecting \(H_{0,j}\) indicates that the three spline terms associated with covariate \(j\) are jointly relevant to the fitted model.

Cross-validation selects \(\widehat\gamma_{\mathrm{CV}}>0\) for \(857\) of the \(1{,}378\) environment pairs. For the remaining \(521\) pairs, \(\widehat\gamma_{\mathrm{CV}}=0\) and the causal-regularized estimator
coincides with OLS. We evaluate differences between the two estimators over the \(857\) pairs with strictly positive regularization.
For each pair of environments, Table~\ref{tab:chamber-results} reports classification metrics to assess the extent to which the rejection pattern recovers the known causal structure. In this analysis, a true positive (TP) is a known direct cause for which the null hypothesis is rejected, whereas a false negative (FN) is a direct cause for which it is not rejected. A false positive (FP) is a covariate that is not a direct cause but is nevertheless rejected, while a true negative (TN) is a non-direct cause for which the null hypothesis is not rejected. 
\begin{table}[t]
\renewcommand{\arraystretch}{1.2}
\centering
\caption{Classification performance for the quadratic spline model with
one knot at zero. The OLS and CR columns report average metrics over the
\(857\) environment pairs for which
\(\widehat{\gamma}_{\mathrm{CV}}>0\). The final column reports averages
over the \(521\) pairs for which
\(\widehat{\gamma}_{\mathrm{CV}}=0\).}
\label{tab:chamber-results}
\begin{tabular}{lccc}
\toprule
& \multicolumn{2}{c}{
  \(\widehat{\gamma}_{\mathrm{CV}}>0\)
}
& \(\widehat{\gamma}_{\mathrm{CV}}=0\) \\
\cmidrule(lr){2-3}
\cmidrule(l){4-4}
\textbf{Metric}
& \textbf{OLS}
& \textbf{CR}
& \textbf{OLS = CR} \\
\midrule
Sensitivity
& \(0.647\) & \(0.637\) & \(0.645\) \\
Specificity
& \(0.641\) & \(0.681\) & \(0.635\) \\
Precision
& \(0.440\) & \(0.475\) & \(0.432\) \\
F1 score
& \(0.518\) & \(0.535\) & \(0.513\) \\
Balanced accuracy
& \(0.644\) & \(0.659\) & \(0.640\) \\
MCC
& \(0.269\) & \(0.305\) & \(0.261\) \\
\bottomrule
\end{tabular}
\end{table}
Across the 857 pairs with strictly positive regularization, two-sample t-tests indicate that sensitivity does not differ significantly different between OLS and causal regularization (p$=0.095$), whereas specificity is significantly higher under CR (p$ = 1.184 \times  10^{-12}$). 
On average, causal regularization produces a more selective rejection pattern, at the cost of a small reduction in sensitivity. Specifically, OLS identifies an average of \(3.23\) of the five direct causes, whereas causal regularization identifies \(3.19\) on average.
For the remaining pairs with
\(\widehat{\gamma}_{\mathrm{CV}}=0\), both sensitivity
and specificity remain close to the corresponding OLS values observed among the pairs with \(\widehat{\gamma}_{\mathrm{CV}}>0\). 

Overall, the modest difference between OLS and causal regularization for this illustration can be understood in terms of the known causal graph. Since \(I_2\) has no children, its Markov blanket consists of its five direct causes, allowing OLS to
capture much of the relevant structure. In this setting, the main benefit of causal regularization is a reduction in false positives rather than an increase in the number of detected direct causes.
\section{Conclusion}
We proposed an extension of causal regularization \citep{kania2025causalregularizationtradeoffinsample} that builds robust prediction models from two heterogeneous environments without requiring either environment to be observational or the underlying data-generating process to be linear. The new mathematical framework distinguishes between solutions that provide stable predictions and causal ones.
In the general robust setting, interventions can affect directly both the input variables and the target outcome. The resulting parameter is interpreted as a stable predictive solution but does not necessarily represent a causal effect. However, under the additional assumptions of mean invariance and inner-product invariance \citep{rothenhausler2019causal}, the causal parameter becomes a stationary solution of the risk difference between-environments and emerges as the population limit of the regularization path. The causal formulation is a particular version of the broader robust framework. 

There are multiple paths for further study based on the results. The current version uses two environments. Generalizing the method to multiple environments \citep{kennerberg2023optimal} could make it possible to exploit richer forms of heterogeneity and to identify which shifts are most informative for robust or causal prediction. Another promising direction is to extend the approach to functional data \citep{kennerberg2023optimal}.

\bibliographystyle{chicago}
\bibliography{biblio}
\appendix

\clearpage

\section{Characterization of \texorpdfstring{\(\beta^\star\)}{beta star} as a Stationary Point}
\label{appendix:betacpstatpoint}

\begin{proof}

Recall that \[ \mathcal{B} = \left\{ \beta \in \mathbb{R}^p : \nabla_\beta R_\Delta(\beta)=0 \right\}. \]
The risk difference can be written as
\[
R_\Delta(\beta)
=
S_\Delta - 2\beta^t Z_\Delta + \beta^t G_\Delta \beta,
\]
Differentiating \(R_\Delta(\beta)\) with respect to \(\beta\) gives
\[
\nabla_\beta R_\Delta(\beta)
=
-2Z_\Delta + 2G_\Delta\beta.
\]
Let \(\beta^\star \in \mathcal{B}\). By definition of \(\mathcal{B}\), we have 
\[
-2Z_\Delta + 2G_\Delta\beta^\star = 0,
\]
which is equivalent to
\[
G_\Delta\beta^\star = Z_\Delta.
\]
Every \(\beta^\star \in \mathcal{B}\) satisfies the stated linear system.
If \(G_\Delta\) is full rank, then it is invertible and the linear system admits the unique solution
\[
\beta^\star = G_\Delta^{-1}Z_\Delta.
\]
\end{proof}

\section{Stationary Solutions}
\label{appendix:kernel}

\begin{proof}
We first show that $G_\Delta$ and $G_\Delta^+$ have the same kernel. Let $G_\Delta=V\Lambda V^t$ be the spectral decomposition of $G_\Delta$, so that
$G_\Delta^+=V|\Lambda|V^t$. For any $x\in\mathbb R^p$,
\[
G_\Delta x=0
\iff \Lambda V^t x=0
\iff |\Lambda|V^t x=0
\iff G_\Delta^+x=0,
\]
where the middle equivalence follows from $\lambda_i=0$ if and only if $|\lambda_i|=0$. It follows that $\ker(G_\Delta)=\ker(G_\Delta^+)$.

We now use this equality to compare two reference solutions. Let $\beta^{1,\star},\beta^{2,\star}\in\mathcal B$. Since both satisfy the stationary equation,
\[
G_\Delta\beta^{1,\star}
=Z_\Delta
=G_\Delta\beta^{2,\star}.
\]
Subtracting gives $G_\Delta(\beta^{1,\star}-\beta^{2,\star})=0$. Their difference belongs to $\ker(G_\Delta)$ and, by equality of the kernels, also to $\ker(G_\Delta^+)$.
Consequently,
\[
G_\Delta^+(\beta^{1,\star}-\beta^{2,\star})=0,
\qquad\text{so}\qquad
G_\Delta^+\beta^{1,\star}
=G_\Delta^+\beta^{2,\star}.
\]
It follows that $Z_\Delta^+=G_\Delta^+\beta^\star$ does not depend on which $\beta^\star\in\mathcal B$ is chosen.

To prove the corresponding result for $S_\Delta^+$, define $d=\beta^{1,\star}-\beta^{2,\star}$. Since $G_\Delta^+d=0$ and $G_\Delta^+$ is symmetric, we also have $d^tG_\Delta^+=0$. Expanding gives 
\[
\begin{aligned}
(\beta^{1,\star})^tG_\Delta^+\beta^{1,\star}
&=(\beta^{2,\star}+d)^tG_\Delta^+
  (\beta^{2,\star}+d)\\
&=(\beta^{2,\star})^tG_\Delta^+\beta^{2,\star}
  +2d^tG_\Delta^+\beta^{2,\star}
  +d^tG_\Delta^+d.
\end{aligned}
\]
The last two terms vanish, so
\[
(\beta^{1,\star})^tG_\Delta^+\beta^{1,\star}
=(\beta^{2,\star})^tG_\Delta^+\beta^{2,\star}.
\]
$S_\Delta^+$ also does not depend on the chosen reference solution.

Finally, for any $\beta\in\mathbb R^p$, we have $\beta-\beta^{1,\star}=\beta-\beta^{2,\star}-d$.
Therefore,
\[
\begin{aligned}
(\beta-\beta^{1,\star})^tG_\Delta^+
(\beta-\beta^{1,\star})
&=(\beta-\beta^{2,\star}-d)^tG_\Delta^+
  (\beta-\beta^{2,\star}-d)\\
&=(\beta-\beta^{2,\star})^tG_\Delta^+
  (\beta-\beta^{2,\star})\\
&\quad-2d^tG_\Delta^+(\beta-\beta^{2,\star})
  +d^tG_\Delta^+d.
\end{aligned}
\]
Again, the last two terms vanish. Thus,
\[
(\beta-\beta^{1,\star})^tG_\Delta^+
(\beta-\beta^{1,\star})
=
(\beta-\beta^{2,\star})^tG_\Delta^+
(\beta-\beta^{2,\star}),
\]
which proves that $R_\Delta^+(\beta)$ does not depend
on which $\beta^\star\in\mathcal B$ is chosen.
\end{proof}

\section{Worst-Case Risk Representation over the Robust Solution Set}
\label{appendix:worst_case}
\begin{proof}

Let
\[
B=\{\beta^\star\in\mathbb R^p: G_\Delta \beta^\star=Z_\Delta\}.
\]
Define
\[
\mathcal C_{\gamma}
=
\left\{
P_{e}:
M_{e}\preceq
\frac12 M_+
+
\frac{\gamma}{2}M_{\Delta}^+
\right\},
\]
where
\[
M_{\Delta}^+
=
\begin{bmatrix}
G_\Delta^+ & Z_{\Delta}^+\\
(Z_{\Delta}^+)^t & S_{\Delta}^+
\end{bmatrix},
\]
with
\[
Z_{\Delta}^+
=
G_\Delta^+\beta^\star,
\qquad
S_{\Delta}^+
=
(\beta^\star)^t G_\Delta^+\beta^\star.
\]
For every distribution \(P_{e}^{X,Y}\in\mathcal C_\gamma\), by definition,
\[
M_{e}
\preceq
M_\gamma=
\frac12 M_+
+
\frac{\gamma}{2}M_{\Delta}^+.
\]
Since
\[
R_{e}(\beta)
=
\begin{bmatrix}
\beta\\
-1
\end{bmatrix}^t
M_{e}
\begin{bmatrix}
\beta\\
-1
\end{bmatrix},
\]
it follows that
\[
R_{e}(\beta)
\le
\begin{bmatrix}
\beta\\
-1
\end{bmatrix}^t
\left(
\frac12 M_+
+
\frac{\gamma}{2}M_{\Delta}^+
\right)
\begin{bmatrix}
\beta\\
-1
\end{bmatrix}.
\]
Expanding the right-hand side gives
\[
R_{e}(\beta)
\le
\frac12 R_+(\beta)
+
\frac{\gamma}{2}
\left[
S_{\Delta}^+
-2\beta^t Z_{\Delta}^+
+\beta^t G_\Delta^+\beta
\right].
\]
Equivalently, we can write
\[
R_{e}(\beta)
\le
\frac12 R_+(\beta)
+
\frac{\gamma}{2}
(\beta-\beta^\star)^t
G_\Delta^+
(\beta-\beta^\star).
\]
Taking the supremum over \(P_{e}^{X,Y}\in\mathcal{U}\cup\mathcal C_\gamma\) gives
\[
\sup_
{P_{X,Y}^{e}\in\mathcal C_{\gamma}}
R_{e}(\beta)
\le
\frac12 R_+(\beta)
+
\frac{\gamma}{2}
R_{\Delta}^+(\beta).
\]
The upper bound is attained by
\(P_{X,Y}^{e}=\mathcal{N}(0,M_\gamma)
\in\mathcal{C}_{\gamma}\), whose second-moment matrix
equals \(M_\gamma\)
\[
\sup_{P_{e}^{X,Y}\in\mathcal C_\gamma}
R_{e}(\beta)
=
\frac{1}{2}R_+(\beta)
+
\frac{\gamma}{2}R_\Delta^+(\beta).
\]

\end{proof}

\section{Closed-form solution of the robust estimator }
\label{appendix:beta_gamma}
\begin{proof}
Recall that 
\[
\beta_\gamma=\arg\min_{\beta\in\mathbb R^p}
\left\{\frac12R_+(\beta)+\frac{\gamma}{2}R_\Delta^+(\beta)\right\}.
\]
Using
\[
R_+(\beta)=S_+-2\beta^tZ_++\beta^tG_+\beta,
\qquad
R_\Delta^+(\beta)
=S_\Delta^+-2\beta^tZ_\Delta^+
+\beta^tG_\Delta^+\beta,
\]
differentiating the objective with respect to $\beta$ and setting the derivative equal to zero yields
\[
0=
\left.
\frac{\partial}{\partial\beta}
\left\{\frac12R_+(\beta)+\frac{\gamma}{2}R_\Delta^+(\beta)\right\}
\right|_{\beta=\beta_\gamma}
=
G_+\beta_\gamma-Z_+
+\gamma(G_\Delta^+\beta_\gamma-Z_\Delta^+).
\]
Since $G_++\gamma G_\Delta^+$ is positive definite,
the objective is strictly convex and its unique minimizer is
\[
\beta_\gamma
=(G_++\gamma G_\Delta^+)^{-1}(Z_++\gamma Z_\Delta^+)
=G_\gamma^{-1}Z_\gamma.
\]
\end{proof}
\section{Consistency of the finite-sample estimator}
\label{appendix:consistency}

\begin{proof}

By the weak law of large numbers,
\[
\mathbb{G}_e\xrightarrow{p}G_e,
\qquad
\mathbb{Z}_e\xrightarrow{p}Z_e.
\]
It follows that
\begin{equation}
\mathbb{G}_+
=
\mathbb{G}_{e_1}+\mathbb{G}_{e_2}
\xrightarrow{p}
G_{e_1}+G_{e_2}
=
G_+,
\label{eq:hat_g+}
\end{equation}
and
\[
\mathbb{G}_\Delta
=
\mathbb{G}_{e_1}-\mathbb{G}_{e_2}
\xrightarrow{p}
G_{e_1}-G_{e_2}
=
G_\Delta.
\]
Similarly,
\[
\mathbb{Z}_+\xrightarrow{p}Z_+,
\qquad
\mathbb{Z}_\Delta\xrightarrow{p}Z_\Delta.
\]
Since replacing the eigenvalues of a symmetric matrix by their absolute values is a continuous operation, the continuous mapping theorem gives
\begin{equation}
\mathbb{G}_\Delta^+
\xrightarrow{p}
G_\Delta^+.
\label{eq:hat_gdelta+}
\end{equation}
By definition,
\[
\mathbb{G}_\gamma
=
\frac{1}{2}\mathbb{G}_+
+
\frac{\gamma}{2}\mathbb{G}_\Delta^+.
\]
Using the convergence results in
Equations~\eqref{eq:hat_g+} and \eqref{eq:hat_gdelta+},
\[
\mathbb{G}_\gamma
\xrightarrow{p}
\frac{1}{2}G_+
+
\frac{\gamma}{2} G_\Delta^+
=
G_\gamma.
\]
The empirical linear term is
\[
\mathbb{Z}_{\gamma}
=
\frac{1}{2}\mathbb{Z}_+
+
\frac{\gamma}{2}\mathbb{G}_\Delta^+\widehat\beta^\star.
\]

Choose \(\beta^\star \in \mathcal{B}\) that solves \(G_\Delta \beta^\star= Z_\Delta\). By spectral decomposition, \begin{equation}
\mathbb{G}_\Delta^+\mathbb{G}_\Delta^\dagger\mathbb{G}_\Delta=\mathbb{G}_\Delta^{+}.
\label{eq:spect_decm}
\end{equation}
Moreover, the product \(\mathbb{G}_\Delta^+\mathbb{G}_\Delta^\dagger\) has eigenvalues in \(\{-1,0,1\}\), so its operator norm is at most one. 

Using Equation \ref{eq:spect_decm} and \(\widehat\beta^\star=\mathbb{G}_\Delta^\dagger \mathbb{Z}_\Delta\), we obtain 

\[
\begin{aligned}
\mathbb G_\Delta^+\widehat\beta^\star
-\mathbb G_\Delta^+\beta^\star
&=\mathbb G_\Delta^+\mathbb G_\Delta^\dagger\mathbb Z_\Delta
-\mathbb G_\Delta^+\mathbb G_\Delta^\dagger
\mathbb G_\Delta\beta^\star\\
&=\mathbb G_\Delta^+\mathbb G_\Delta^\dagger
(\mathbb Z_\Delta-\mathbb G_\Delta\beta^\star).
\end{aligned}
\]
Since the product $\mathbb G_\Delta^+\mathbb G_\Delta^\dagger$
has operator norm at most one,
\[
\left\|\mathbb G_\Delta^+\widehat\beta^\star
-\mathbb G_\Delta^+\beta^\star\right\|
\leq
\left\|\mathbb Z_\Delta-\mathbb G_\Delta\beta^\star\right\|
\xrightarrow{p}0,
\]
where the convergence follows from
$G_\Delta\beta^\star=Z_\Delta$ and convergence of the
empirical moments. Together with Equation \ref{eq:hat_gdelta+}, this implies
\[
\mathbb G_\Delta^+\widehat\beta^\star
\xrightarrow{p}G_\Delta^+\beta^\star=Z_\Delta^+.
\]

Consequently, for fixed $\gamma\geq0$,
\[
\mathbb Z_\gamma
=\frac{1}{2}\mathbb Z_++\frac{\gamma}{2}\mathbb G_\Delta^+\widehat\beta^\star
\xrightarrow{p}\frac{1}{2}Z_++\frac{\gamma}{2} Z_\Delta^+=Z_\gamma.
\]

Since $G_\gamma$ is positive definite, the Moore--Penrose
pseudoinverse is continuous at the limit. By the continuous mapping theorem, it follows that
\[
\widehat\beta_\gamma
=\mathbb G_\gamma^\dagger\mathbb Z_\gamma
\xrightarrow{p}
(G_\gamma)^{-1}(Z_\gamma)
=\beta_\gamma.
\]

\end{proof}

\section{Conditional variance of the finite sample estimator}
\label{appendix:conditional-variance}

For a fixed value of \(\gamma\geq 0\), conditional on \(\mathbf X\), the matrix \(\mathbb{G}_\gamma^{\dagger}\) is non random and symmetric. 
It follows that

\begin{equation}
\operatorname{Var}\!\left(\widehat\beta_\gamma\mid \mathbf X\right)
=
\mathbb G_\gamma^{\dagger} \operatorname{Var}(\mathbb{Z}_\gamma\mid \mathbf X )\mathbb G_\gamma^{\dagger}. 
\label{eq:var_beta}
\end{equation}

The conditional variance of \(\mathbb Z_\gamma\) is derived below. Conditioning on $\mathbf{X}$ means conditioning on both \((\mathbf X)\).

\begin{proof}

By definition,
\[
\mathbb Z_\gamma
=
\frac{1}{2}\mathbb Z_+
+
\frac{\gamma}{2}\mathbb Z_\Delta^+,
\]
expanding its conditional variance gives
\begin{equation}
\begin{aligned}
\operatorname{Var}(\mathbb{Z}_\gamma\mid \mathbf X)
&=\operatorname{Var}\left(\frac{1}{2}\mathbb{Z}_+\mid \mathbf X\right)
+
\operatorname{Var}\left(\frac{\gamma}{2}\mathbb{Z}_\Delta^+\mid \mathbf X\right)
                  +
                \frac{\gamma}{2}
\operatorname{Cov}(\mathbb Z_\Delta^+,\mathbb Z_+
                  \mid \mathbf X)
\end{aligned}
\label{eq:variance}
\end{equation}

For the first term, independence between environments gives
\[
\begin{aligned}
\operatorname{Var}\left(\frac12\mathbb Z_+\mid\mathbf X\right)
&=
\frac14\operatorname{Var}\left(
\frac{\mathbf X_{e_1}^t\mathbf Y_{e_1}}{n_{e_1}}
+
\frac{\mathbf X_{e_2}^t\mathbf Y_{e_2}}{n_{e_2}}
\;\middle|\;\mathbf X
\right)\\
&=
\frac14\left[
\frac{
\mathbf X_{e_1}^t
\operatorname{Var}(\mathbf Y_{e_1}\mid\mathbf X_{e_1})
\mathbf X_{e_1}
}{n_{e_1}^2}
+
\frac{
\mathbf X_{e_2}^t
\operatorname{Var}(\mathbf Y_{e_2}\mid\mathbf X_{e_2})
\mathbf X_{e_2}
}{n_{e_2}^2}
\right]\\
&=\frac14\Sigma_+.
\end{aligned}
\]

For the second term, using
$\widehat\beta^\star
=\mathbb G_\Delta^\dagger\mathbb Z_\Delta$
and
$\mathbb Z_\Delta^+
=\mathbb G_\Delta^+\widehat\beta^\star$,
we obtain
\[
\begin{aligned}
\operatorname{Var}\left(
\frac{\gamma}{2}\mathbb Z_\Delta^+\mid\mathbf X
\right)
&=
\frac{\gamma^2}{4}
\operatorname{Var}\left(
\mathbb G_\Delta^+\mathbb G_\Delta^\dagger
\mathbb Z_\Delta
\mid\mathbf X
\right)\\
&=
\frac{\gamma^2}{4}
\mathbb G_\Delta^+\mathbb G_\Delta^\dagger
\operatorname{Var}\left(
\frac{\mathbf X_{e_1}^t\mathbf Y_{e_1}}{n_{e_1}}
-
\frac{\mathbf X_{e_2}^t\mathbf Y_{e_2}}{n_{e_2}}
\;\middle|\;\mathbf X
\right)
\mathbb G_\Delta^\dagger\mathbb G_\Delta^+\\
&=
\frac{\gamma^2}{4}
\mathbb G_\Delta^+\mathbb G_\Delta^\dagger
\Sigma_+
\mathbb G_\Delta^\dagger\mathbb G_\Delta^+.
\end{aligned}
\]

For the cross-covariance terms, define the corresponding difference
\[
\Sigma_\Delta
=
\frac{
\mathbf X_{e_1}^t
\operatorname{Var}(\mathbf Y_{e_1}\mid\mathbf X_{e_1})
\mathbf X_{e_1}
}{n_{e_1}^2}
-
\frac{
\mathbf X_{e_2}^t
\operatorname{Var}(\mathbf Y_{e_2}\mid\mathbf X_{e_2})
\mathbf X_{e_2}
}{n_{e_2}^2}.
\]
Then
\[
\begin{aligned}
\operatorname{Cov}\left(
\frac12\mathbb Z_+,
\frac{\gamma}{2}\mathbb Z_\Delta^+
\mid\mathbf X
\right)&=
\frac{\gamma}{4}
\operatorname{Cov}\left(
\frac{\mathbf X_{e_1}^t\mathbf Y_{e_1}}{n_{e_1}}
+
\frac{\mathbf X_{e_2}^t\mathbf Y_{e_2}}{n_{e_2}},
\frac{\mathbf X_{e_1}^t\mathbf Y_{e_1}}{n_{e_1}}
-
\frac{\mathbf X_{e_2}^t\mathbf Y_{e_2}}{n_{e_2}}
\;\middle|\;\mathbf X
\right)
\mathbb G_\Delta^\dagger\mathbb G_\Delta^+\\
&=
\frac{\gamma}{4}
\Sigma_\Delta
\mathbb G_\Delta^\dagger\mathbb G_\Delta^+.
\end{aligned}
\]

Substituting these expressions into
Equation~\ref{eq:variance}
gives the conditional variance of $\mathbb Z_\gamma$

\begin{equation}
\begin{aligned}
\operatorname{Var}(\mathbb Z_\gamma\mid\mathbf X)
={}&
\frac14\Sigma_+
+
\frac{\gamma^2}{4}
\mathbb G_\Delta^+\mathbb G_\Delta^\dagger
\Sigma_+
\mathbb G_\Delta^\dagger\mathbb G_\Delta^+\\
&+
\frac{\gamma}{4}\left(
\mathbb G_\Delta^+\mathbb G_\Delta^\dagger\Sigma_\Delta
+ \Sigma_\Delta\mathbb G_\Delta^\dagger\mathbb G_\Delta^+
\right).
\end{aligned}
\label{eq:variance_complete}
\end{equation}

Finally, by plugging Equation \ref{eq:variance_complete} in Equation \ref{eq:var_beta}
we obtain the conditional variance of $\widehat{\beta}_\gamma$

\[
\begin{aligned}
\operatorname{Var}(\widehat\beta_\gamma\mid\mathbf X)
&=
\mathbb G_\gamma^\dagger
\operatorname{Var}(\mathbb Z_\gamma\mid\mathbf X)
\mathbb G_\gamma^\dagger\\
&=
\frac14\mathbb G_\gamma^\dagger
\Bigl[
(I+\gamma\mathbb G_\Delta^+\mathbb G_\Delta^\dagger)
\Sigma_+
(I+\gamma\mathbb G_\Delta^\dagger\mathbb G_\Delta^+)\\
&\quad+
\gamma\left(
\mathbb G_\Delta^+\mathbb G_\Delta^\dagger\Sigma_\Delta
+
\Sigma_\Delta\mathbb G_\Delta^\dagger\mathbb G_\Delta^+
\right)\\
&\quad-
\gamma\left(
\mathbb G_\Delta^+\mathbb G_\Delta^\dagger\Sigma_+
+
\Sigma_+\mathbb G_\Delta^\dagger\mathbb G_\Delta^+
\right)
\Bigr]
\mathbb G_\gamma^\dagger.
\end{aligned}
\]
\end{proof}

\section{Causal limit}
\label{appendix:causal_theorem}
\begin{proof}
By definition,
\[
r_{CP}^e
=
Y^e-(X^e)^t\beta_{CP}.
\]
Multiplying by \(X^e\) and taking expectations gives
\[
\mathbb E[X^e r_{CP}^e]
=
\mathbb E[X^eY^e]
-
\mathbb E[X^e(X^e)^t]\beta_{CP}.
\]
Using
\[
G^e=\mathbb E[X^e(X^e)^t],
\qquad
Z^e=\mathbb E[X^eY^e],
\]
we obtain
\[
\mathbb E[X^e r_{CP}^e]
=
Z^e-G^e\beta_{CP}.
\]
By inner-product invariance, this quantity is constant across environments.
For any two environments \(e_1,e_2\) such that
\(
P_{e_1}^{X,Y},P_{e_2}^{X,Y}\in \mathcal{P_I},
\)
\[
Z^{e_1}-G^{e_1}\beta_{CP}
=
Z^{e_2}-G^{e_2}\beta_{CP}.
\]
Rearranging gives
\[
(G^{e_2}-G^{e_1})\beta_{CP}
=
Z^{e_2}-Z^{e_1}.
\]

Equivalenty,
\(
G_\Delta\beta_{CP}=Z_\Delta
\). 
Under the theorem's assumption that $G_\Delta$ is nonsingular, this solution is unique, so $\beta^\star=\beta_{\mathrm{CP}}$.

To establish the regularization limit, note that $G_\Delta^+$ is positive definite. For $\gamma>0$, factoring out $\gamma/2$ from both $G_\gamma$ and $Z_\gamma$ gives
\[
\beta_\gamma
=
\left(\frac{1}{\gamma}G_++G_\Delta^+\right)^{-1}
\left(\frac{1}{\gamma}Z_++Z_\Delta^+\right).
\]
Taking the limit as $\gamma\to\infty$ and using $Z_\Delta^+=G_\Delta^+\beta_{\mathrm{CP}}$, we obtain
\[
\beta_{\mathrm{RL}}
=\lim_{\gamma\to\infty}\beta_\gamma
=(G_\Delta^+)^{-1}Z_\Delta^+
=\beta_{\mathrm{CP}}.
\]

\end{proof}
\section{Limit of anchor regression with a binary anchor}
\label{appendix:ar}
We derive the limiting anchor-regression estimator for a centered binary environment anchor. We consider the model without the intercept and allow the two environments to have different sample sizes.

The centered binary anchor is
\[
\begin{aligned}
\widetilde{\mathbf{D}}
&=
\mathbf D-\bar D\mathbf 1_n\\
&=
\mathbf D-\frac{n_2}{n}\mathbf 1_n\\
&=
\begin{bmatrix}
-\dfrac{n_2}{n}\mathbf 1_{n_1}\\[2mm]
 \dfrac{n_1}{n}\mathbf 1_{n_2}
\end{bmatrix}.
\end{aligned}
\]

The orthogonal projection onto the space spanned by \(\widetilde{\mathbf{D}}\) is
\[
 \mathbf{P}_{\widetilde{\mathbf{D}}}
=
\widetilde{\mathbf{D}}
\left(\widetilde{\mathbf{D}}^t\widetilde{\mathbf{D}}\right)^{-1}
\widetilde{\mathbf{D}}^t.
\]
For a candidate coefficient vector \(\beta\in\mathbb{R}^p\), define the
pooled residual vector by
\[
\mathbf{r}(\beta)
=
\mathbf y- \mathbf X\beta
=
\begin{bmatrix}
\mathbf{r}^{e_1}(\beta)\\
\mathbf{r}^{e_2}(\beta)
\end{bmatrix},
\qquad
r_i(\beta)=Y_i-X_i^t \beta.
\]

Firstly, we calculate the squared norm of the centered anchor:
\[
\begin{aligned}
\widetilde{\mathbf{D}}^t\widetilde{\mathbf{D}}
&=
\begin{bmatrix}
-\dfrac{n_2}{n}\mathbf{1}_{n_1}^t
&
\phantom{-}\dfrac{n_1}{n}\mathbf{1}_{n_2}^t
\end{bmatrix}
\begin{bmatrix}
-\dfrac{n_2}{n}\mathbf{1}_{n_1}\\[3mm]
\phantom{-}\dfrac{n_1}{n}\mathbf{1}_{n_2}
\end{bmatrix}\\
&=
\left(
-\frac{n_2}{n}\mathbf{1}_{n_1}^t
\right)
\left(
-\frac{n_2}{n}\mathbf{1}_{n_1}
\right)
+
\left(
\frac{n_1}{n}\mathbf{1}_{n_2}^t
\right)
\left(
\frac{n_1}{n}\mathbf{1}_{n_2}
\right)\\
&=
\frac{n_2^2}{n^2}
\mathbf{1}_{n_1}^t\mathbf{1}_{n_1}
+
\frac{n_1^2}{n^2}
\mathbf{1}_{n_2}^t\mathbf{1}_{n_2}\\
&=
\frac{n_1n_2^2}{n^2}
+
\frac{n_2n_1^2}{n^2}\\
&=
\frac{n_1n_2}{n}.
\end{aligned}
\]
Next, using the matrix form of \(\widetilde{\mathbf{D}}\), we obtain

\[
\begin{aligned}
\widetilde{\mathbf{D}}^t \mathbf{r}(\beta)
&=
\begin{bmatrix}
-\dfrac{n_2}{n}\mathbf{1}_{n_1}^t
&
\phantom{-}\dfrac{n_1}{n}\mathbf{1}_{n_2}^t
\end{bmatrix}
\begin{bmatrix}
\mathbf{r}^{e_1}(\beta)\\
\mathbf{r}^{e_2}(\beta)
\end{bmatrix}\\[2mm]
&=
-\frac{n_2}{n}
\mathbf{1}_{n_1}^t \mathbf{r}^{e_1}(\beta)
+
\frac{n_1}{n}
\mathbf{1}_{n_2}^t \mathbf{r}^{e_2}(\beta)\\
&=
-\frac{n_2}{n}
n_1\overline r_{e_1}(\beta)
+
\frac{n_1}{n}
n_2\overline r_{e_2}(\beta)\\
&=
\frac{n_1n_2}{n}
\left(
\overline r_{e_2}(\beta)-\overline r_{e_1}(\beta)
\right).
\end{aligned}
\]
where \(\overline r_{e_1}(\beta),\quad \overline r_{e_2}(\beta)\) are the average residual in each environment.
The projection of the residual onto the anchor space can be written as
\[
\begin{aligned}
\mathbf P_{\widetilde{\mathbf{D}}}\mathbf{r}(\beta)
&=
\widetilde{\mathbf{D}}
\left(\widetilde{\mathbf{D}}^t\widetilde{\mathbf{D}}\right)^{-1}
\widetilde{\mathbf{D}}^t \mathbf{r}(\beta)\\
&=
\widetilde{\mathbf{D}}
\left(
\bar r_{e_2}(\beta)-\bar r_{e_1}(\beta)
\right).
\end{aligned}
\]

Taking the squared euclidean norm
\begin{equation}
\begin{aligned}
\left\|\mathbf P_{\widetilde{\mathbf{D}}}\mathbf{r}(\beta)\right\|_2^2
&=
\left(
\bar r_{e_2}(\beta)-\bar r_{e_1}(\beta)
\right)^2
\widetilde{\mathbf{D}}^t\widetilde{\mathbf{D}}\\
&=
\frac{n_1n_2}{n}
\left(
\bar r_{e_2}(\beta)-\bar r_{e_1}(\beta)
\right)^2.
\end{aligned}
\label{eq:proj1}
\end{equation}

The anchor regression estimator is defined as
\[
\widehat\beta_{\mathrm{AR}}(\gamma)
=
\arg\min_\beta
\left\{
\frac{1}{n}
\left\|(I-\mathbf P_{\widetilde{\mathbf{D}}})\mathbf{r}(\beta)\right\|_2^2
+
\frac{\gamma}{n}
\left\|\mathbf P_{\widetilde{\mathbf{D}}}\mathbf{r}(\beta)\right\|_2^2
\right\}.
\]

Equivalently, using Equation \eqref{eq:proj1}
\[
\widehat\beta_{\mathrm{AR}}(\gamma)
=
\arg\min_\beta
\left\{
\frac{1}{n}\|\mathbf y - \mathbf X\beta\|_2^2
+
(\gamma-1)\frac{n_1n_2}{n^2}
\left(
\bar r_{e_2}(\beta)-\bar r_{e_1}(\beta)
\right)^2
\right\}.
\]
Anchor regression penalizes differences between the average
residuals in the two environments.

When \(\gamma=1\), the additional penalty relative to ordinary least
squares vanishes:
\[
\widehat\beta_{\mathrm{AR}}(1)
=
\widehat\beta_{\mathrm{OLS}},
\]

As \(\gamma\rightarrow\infty\), the projected residual is increasingly
penalized. Provided that the projected residual can be set equal to zero,
the limiting estimator must satisfy
\[
\widetilde{\mathbf{D}}^t \mathbf{r}(\beta)=\frac{n_1n_2}{n}
\left(
\bar r_{e_2}(\beta)-\bar r_{e_1}(\beta)
\right)=0.
\]

Because \(n_1,n_2>0\), the limiting condition becomes
\[
\bar r_{e_1}(\beta)=\bar r_{e_2}(\beta).
\]

The limiting anchor-regression estimator is the constrained
least-squares estimator
\[
\widehat\beta_{\mathrm{AR},\infty}
=
\arg\min_\beta\|\mathbf{y}-\mathbf X\beta\|_2^2
\quad\text{subject to}\quad
\widetilde{\mathbf{D}}^t( \mathbf y-\mathbf X\beta)=0.
\]

The Lagrangian associated with the constrained problem is
\[
\mathcal L(\beta,\lambda)
=
(\mathbf y- \mathbf X\beta)^t(\mathbf y-\mathbf X\beta)
+
2\lambda
\left(
\widetilde{\mathbf{D}}^t \mathbf X\beta-\widetilde{\mathbf{D}}^t \mathbf y
\right).
\]

Differentiating with respect to \(\beta\) gives
\[
\frac{\partial\mathcal L(\beta,\lambda)}{\partial \beta}
=
-2\mathbf X^t(\mathbf y-\mathbf X\beta)
+
2\lambda \mathbf X^t\widetilde{\mathbf{D}}.
\]
The first-order condition is 
\[
-2\mathbf X^t(\mathbf y-\mathbf X\beta)
+
2\lambda \mathbf X^t\widetilde{\mathbf{D}}
=
0.
\]
Dividing by \(2\) and rearranging yields

\begin{equation}
\beta=
\widehat\beta_{\mathrm{OLS}}
-
\lambda
(\mathbf X^t \mathbf X)^{-1}\mathbf X^t\widetilde{\mathbf{D}}.
\label{eq:limit_sol_AR}
\end{equation}

Differentiating the Lagrangian with respect to \(\lambda\) recovers the
constraint:
\[
\frac{\partial\mathcal L(\beta,\lambda)}{\partial\lambda}
=
2\left(
\widetilde{\mathbf{D}}^t \mathbf X\beta-\widetilde{\mathbf{D}}^t \mathbf y
\right)
=
0.
\]
Substituting Equation \ref{eq:limit_sol_AR} into this constraint gives
\[
\begin{aligned}
0
&=
\widetilde{\mathbf{D}}^t \mathbf X
\left[
\widehat\beta_{\mathrm{OLS}}
-
\lambda
(\mathbf X^t \mathbf X)^{-1} \mathbf X^t\widetilde{\mathbf{D}}
\right]
-
\widetilde{\mathbf{D}}^t \mathbf y\\
&=
\widetilde{\mathbf{D}}^t \mathbf X \widehat\beta_{\mathrm{OLS}}
-
\lambda
\widetilde{\mathbf{D}}^t \mathbf X
(\mathbf X^t \mathbf X)^{-1}
\mathbf X^t\widetilde{\mathbf{D}}
-
\widetilde{\mathbf{D}}^t \mathbf y.
\end{aligned}
\]
It follows that
\[
\lambda
\widetilde{\mathbf{D}}^t \mathbf X
(\mathbf X^t \mathbf X)^{-1}
\mathbf X^t\widetilde{\mathbf{D}}
=
\widetilde{\mathbf{D}}^t \mathbf X\widehat\beta_{\mathrm{OLS}}
-
\widetilde{\mathbf{D}}^t \mathbf y 
\]
\begin{equation}
\lambda
=
\frac{
\widetilde{\mathbf{D}}^t \mathbf X\widehat\beta_{\mathrm{OLS}}
-
\widetilde{\mathbf{D}}^t \mathbf y
}{
\widetilde{\mathbf{D}}^t \mathbf X
(\mathbf X^t \mathbf X)^{-1}
\mathbf X^t\widetilde{\mathbf{D}}
}.
\label{eq:lambda_eq}
\end{equation}

Substituting Equation \ref{eq:lambda_eq} into Equation \ref{eq:limit_sol_AR}, gives

\[
\widehat\beta_{\mathrm{AR},\infty}
=
\widehat\beta_{\mathrm{OLS}}
+
(\mathbf X^t \mathbf X)^{-1}\mathbf X^t\widetilde{\mathbf{D}}
\frac{
\widetilde{\mathbf{D}}^t \mathbf y
-
\widetilde{\mathbf{D}}^t X\widehat\beta_{\mathrm{OLS}}
}{
\widetilde{\mathbf{D}}^t \mathbf X
(\mathbf X^t \mathbf X)^{-1}
\mathbf X^t\widetilde{\mathbf{D}}
}.
\]

\end{document}